\documentclass[a4paper,11pt]{article}
\usepackage{jheppub} 
\usepackage{lineno}
\usepackage{graphicx} 
\usepackage{amsmath}
\usepackage{tikz}
\usepackage{caption}
\usepackage{subcaption}
\usepackage{amssymb}
\usepackage{braket}
\usepackage{enumitem}

\title{\boldmath Holographic Casimir Effect for Mixed Boundary Conditions and non-CFTs}

\author{Bing-Ji Ding, Hua-Chao Liu, Rong-Xin Miao}
\affiliation{School of Physics and Astronomy, Sun Yat-sen University,\\
2 Daxue Road, Zhuhai 519082, China}

\emailAdd{miaorx@mail.sysu.edu.cn}

\abstract{This paper investigates the holographic Casimir effect in AdS/BCFT with a brane-localized scalar field. The scalar field takes different values at the strip's two boundaries, leading to mixed boundary conditions. The massive brane-localized scalar field typically breaks the boundary conformal symmetries, allowing us to study the holographic Casimir effect in non-CFTs. In two-dimensional spacetime, we apply the holographic g-theorem to prove the holographic bound on the Casimir effect for general brane-localized matter fields. In higher dimensions, we find that the brane-localized scalar field reduces the Casimir amplitude when the mass squared \(m^2=0\), but can increase it when \(m^2 < 0\). Additionally, there is a no-hair theorem for the cases where \(m^2>0\).
These findings indicate that relevant boundary deformations can enhance the Casimir effect. As a byproduct, we obtain a gravitational dual of the repulsive Casimir force under mixed boundary conditions and argue that it aligns with the cosmic censorship conjecture.}

\begin{document}

\maketitle

\flushbottom

\section{Introduction}

The Casimir energy \cite{Casimir:1948dh} is defined as the lowest energy for a quantum system with fixed boundary conditions or specified background fields \cite{Plunien:1986ca, Bordag:2001qi, Milton:2004ya, Bordag:2009zz}. It has wide applications across condensed matter physics \cite{Plunien:1986ca, Bordag:2001qi, Milton:2004ya, Bordag:2009zz}, cosmology \cite{Li:2009pm, Wang:2016och}, and high-energy physics \cite{Vepstas:1984sw}. In a unitary theory, the energy is expected to be bounded from below. Recently, it was conjectured that the AdS/BCFT \cite{Takayanagi:2011zk} with minimal brane tension establishes a lower bound for the Casimir effect in d-dimensional conformal field theories (CFTs) \cite{Miao:2024gcq, Miao:2025utb}: 
\begin{align}\label{Intro: Casimir bound} 
-\frac{\kappa_1}{C_D} \ge \lim_{T\to T_{\text{min}}} \left(-\frac{\kappa_1}{C_D}\right)|_{\text{holo}}=\frac{-2^{d-2} d \pi^{d-\frac{1}{2}} \Gamma\left(\frac{d-1}{2}\right) \Gamma\left(\frac{1}{d}\right)^d}{\Gamma(d+2) \left(d \Gamma\left(\frac{1}{2}+\frac{1}{d}\right)\right)^{d} }, 
\end{align} 
where \(\kappa_1\) is the Casimir amplitude of a strip, \(C_D\) is the norm of the displacement operator, and \(T_{\text{min}} = -(d-1)\) represents the minimal brane tension (with the AdS radius set to \(l = 1\)). This conjecture has been proven for general 2D CFTs and has been tested in free theories, the Ising model, and \(O(N)\) models for \(N = 2, 3\) in 3D CFTs \cite{Miao:2024gcq, Miao:2025utb}. Remarkably, unlike the Kovtun-Son-Starinets (KSS) bound for fluids \cite{Kovtun:2004de}, this holographic bound for the Casimir effect is universal and independent of the specific gravity theories employed \cite{Miao:2024gcq, Miao:2025utb}, such as Dvali-Gabadadze-Porrati (DGP) gravity \cite{Dvali:2000hr} and Gauss-Bonnet gravity.

While the original work \cite{Miao:2024gcq, Miao:2025utb} mainly focused on CFTs, an important open question is how widely these results can be generalized to non-CFTs. The simplest examples of non-CFTs are massive free theories, which have been shown to adhere to the holographic bound \cite{Miao:2024gcq} \footnote{The norm of the displacement operator is not a constant in non-CFTs. In the case of massive free theories, the bound of \(-\kappa_1/C_D\) holds, provided that we define \(C_D\) in the short-distance limit, specifically \(C_D|_{\text{massive}} = C_D(y \to 0)\). We emphasize that defining the ratio \(-\kappa_1/C_D\) for non-CFTs remains an open question.}. The Casimir effect arises from long-range quantum correlations \cite{Dantchev:2022hvy}. However, the presence of mass shortens these correlations, thereby diminishing the Casimir effect. In this study, we explore another category of non-CFTs by applying matter fields solely to the end-of-the-world (EOW) brane. This approach typically breaks the boundary conformal symmetries while preserving the bulk conformal symmetries. One significant advantage of this configuration is that the AdS soliton still provides a valid bulk solution. The brane-located matter fields only affect the shape of the EOW brane.

We investigate the holographic strip and wedge, which incorporate a brane-localized scalar field \cite{Kanda:2023zse}. The massless brane-localized scalar induces an exact boundary marginal deformation, resulting in a new family of BCFTs. We provide analytical proofs for negative brane tensions and numerical verifications for general brane tensions that these new BCFTs satisfy the holographic bound of the Casimir effect (\ref{Intro: Casimir bound}) for BCFTs. Conversely, massive brane-localized scalars correspond to boundary irrelevant and relevant deformations for \(m^2>0\) and \(m^2<0\), respectively. This generally transforms a CFT into a non-CFT \cite{Kanda:2023zse, Liu:2025gle}. In non-CFTs, the norm of the displacement operator \(C_D\) is no longer constant, raising the open question of how to extend the bound on the ratio \(-\kappa_1/C_D\) to non-CFTs. For simplicity, we focus instead on the Casimir amplitude \(\kappa_1\). It is important to note that \(\kappa_1\) is defined by the Casimir pressure \(T_{nn} = -(d-1)\kappa_1/L^d\), which is an experimentally observable quantity and thus well-defined for non-CFTs.  For \(d=2\), we apply the holographic g-theorem to prove that brane-localized matter fields decrease the Casimir amplitude, provided they satisfy the null energy condition. For \(d \geq 3\), we find that massive brane-localized scalars may increase the Casimir amplitude when \(m^2 < 0\). Additionally, we prove a no-hair theorem for the case where \(m^2 > 0\) and there is zero boundary scalar source. In this scenario, the brane-localized scalar cannot support a connected EOW brane linking the two boundaries of the strip and wedge.

As a byproduct, we obtain a gravity dual of the repulsive Casimir force. To the best of our knowledge, this is the first demonstration of the holographic repulsive Casimir effect. The Casimir force between two planes is typically attractive when the same boundary conditions are applied to both planes \cite{Bachas:2006ti, Diatlyk:2024qpr}. A repulsive Casimir force can only be generated when different boundary conditions are imposed on the two boundaries. In our scenario, the brane-localized scalar field takes different values at each boundary, thereby realizing mixed boundary conditions \cite{Kanda:2023zse} that allow a repulsive Casimir force. According to \cite{Huang:2026liq}, the repulsive Casimir effect is dual to a bulk spacetime that includes a naked singularity in the absence of brane-localized scalars. Fortunately, in our case, the brane-localized scalar bends the EOW brane inward, converting the configuration from disconnected to connected. Consequently, the singularity becomes hidden behind the EOW brane, which aligns with the cosmic censorship hypothesis \cite{Penrose:1969pc, Penrose123}.

The paper is organized as follows: In Section 2, we review the basic theory of the holographic Casimir effect and utilize the holographic g-theorem to prove the holographic bound of the Casimir effect for general brane-localized matter fields in dimension \(d=2\). In Section 3, we analyze the holographic Casimir effect of a strip with massless and massive brane-localized scalar fields. In Section 4, we extend our discussion to the holographic wedge. In Section 5, we explore the gravity dual of the repulsive Casimir force and explain how it aligns with the cosmic censorship conjecture. Finally, in Section 6, we conclude by discussing open issues.

 \begin{figure}[htbp]
  \centering
\includegraphics[width=0.6\textwidth]{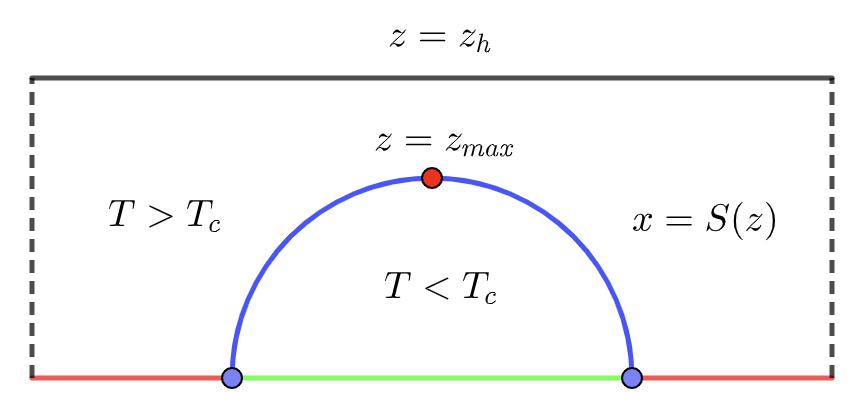}
 \caption{Gravity dual of a strip. The blue points represent the strip boundaries, while the blue curve indicates the EOW brane, i.e., $x=S(z)$. Due to the periodic nature of the bulk spacetime, the two dashed lines are identified. The gravity dual corresponds to the region between the blue curve and the green line for the brane tension \(T < T_c\). Conversely, the complement of this region represents the gravity dual for the brane tension \(T > T_c\).} 
 \label{GravityDual}
\end{figure}

\section{Holographic Casimir effect and g Theorem}

In this section, we start by reviewing the general theories of the holographic Casimir effect. We then explore an interesting connection between the holographic bound of the Casimir effect (\ref{Intro: Casimir bound}) and the holographic g-theorem \cite{Takayanagi:2011zk,Fujita:2011fp}. Our focus is on the gravitational dual of a strip with mixed boundary conditions for both CFTs and non-CFTs, which can be achieved by introducing matter fields on the end-of-the-world (EOW) branes \cite{Kanda:2023zse}. For example, a massless brane-localized scalar field creates an exact boundary marginal deformation, resulting in a new CFT. In contrast, massive brane-localized scalar fields typically disrupt the boundary conformal symmetries, leading to non-CFTs. By applying the holographic g-theorem, we demonstrate that brane-localized matter fields generally reduce the Casimir effect in two dimensions, thus adhering to the holographic bound (\ref{Intro: Casimir bound}) for this dimension. However, in higher dimensions, the holographic g-theorem alone is insufficient to derive the holographic bound of the Casimir effect. Consequently, we will need to examine specific theories in the following section.

\subsection{Holographic Casimir effect}

Let us start with the Casimir effect of a strip for a $d$-dimensional BCFT. Due to the geometric symmetry, Weyl invariance \(\langle T^i_{\ i} \rangle = 0\), and energy conservation \(\langle \partial_i T^{ij} \rangle = 0\), we can determine the vacuum expectation value of the energy-momentum tensor as follows \cite{Miao:2024gcq}:
\begin{align}\label{sect 2: strip Tij}
\langle T^{i}_{\ j} \rangle_{\text{strip}}=\frac{\kappa_1}{L^d} \text{diag}\Big(-(d-1),1,\dots,1 \Big),
\end{align}
where \(L\) is the width of the strip, and \(\kappa_1\) is the Casimir amplitude.  It turns out the AdS soliton can reproduce the above stress tensor, making it a strong candidate for the gravity dual of the strip \cite{Takayanagi:2011zk}. Additional support for this comes from the principle of topological censorship \cite{Witten:1999xp, Galloway:1999br}. Further discussions on the consistency of the AdS soliton with topological censorship can be found in \cite{Huang:2026liq}.

The metric of AdS soliton and the embedding function of EOW brane are given by
\begin{align}\label{sect 2: soliton}
&\text{metric}:\ ds^2=\frac{\frac{dz^2}{f(z)}+f(z)dx^2-dt^2+dy_a^2}{z^2}, \\
&\text{brane}:\ x=S(z), \label{sect 2: brane}
\end{align}
where $f(z)=1-z^d/z_h^d$. The geometry of the holographic strip is illustrated in Fig. \ref{GravityDual}. The blue points represent the two boundaries of the strip, while the blue curve denotes the EOW brane. Due to the periodic nature of the bulk spacetime, the two dashed lines are identified. The holographic strip is defined as the area between the blue curve and the green line for small brane tension (\(T < T_c\)). In contrast, the complement of this region represents the gravitational dual for large brane tension (\(T > T_c\)) \cite{Takayanagi:2011zk}. The critical tension $T_c$ will be determined later. 

To eliminate the conical singularity of the AdS soliton (\ref{sect 2: soliton}), we fix the bulk period of \( x \) as:
\begin{align}\label{sect 2: beta}
\beta=\frac{4\pi}{|f'(z_h)|}=\frac{4 \pi z_h}{d}.
\end{align}
Note that, on the AdS boundary, the range of \( x \) is labeled as \( 0 \le x \le L \) for a strip. By applying the holographic renormalization \cite{deHaro:2000vlm}, we derive the pressure
\begin{align}\label{sect 2: Txx}
T_{xx}=-(d-1)/z_h^d. 
\end{align} 
Comparing this expression with (\ref{sect 2: strip Tij}), we obtain the holographic Casimir amplitude
\begin{eqnarray}\label{sect2: kappa1}
\kappa_1=\frac{L^d}{z_h^d}.
\end{eqnarray}
For simplicity, we set $z_h=1$ in the following discussions. 

The strip width \( L \) can be determined by applying the Neumann boundary condition (NBC) on the EOW brane, assuming \( 16\pi G_N=1 \): 
\begin{align}\label{sect 2: NBC}
K_{ij}-(K-T) h_{ij}=\sigma T_{\text{matter}\ ij},
\end{align}
where \( K_{ij} \) represents the extrinsic curvature, \( T = (d - 1) \tanh(\rho) \) denotes the brane tension, \( h_{ij} \) indicates the induced metric on the EOW brane, and \( T_{\text{matter}\ ij} \) refers to the stress-energy tensors of the brane-localized matter fields. Note that \( \sigma=1 \) corresponds to normal matter, while \( \sigma=-1 \) designates a ghost. For a brane-localized scalar, we have
\begin{align}\label{sect 2: matter Tij}
T_{\text{matter}\ ij}=\partial_i \phi \partial_j \phi -\frac{1}{2} h_{ij}\Big( h^{kl}\partial_k \phi \partial_l \phi +2 V(\phi)\Big).
\end{align}
The equations of motion (EOM) that govern the brane-localized scalar field is given by: 
\begin{align}\label{sect 2: matter EOM}
D^iD_i \phi=\frac{1}{\sqrt{-h}}\partial_i(\sqrt{-h} h^{ij}\partial_j \phi)=V'(\phi) ,
\end{align}
where \( D_i \) is the covariant derivative associated with the induced metric \( h_{ij} \).

First, consider the case where the brane tension is below the critical value. From the NBC (\ref{sect 2: NBC}) and the EOM (\ref{sect 2: matter EOM}), we can determine \( S'(z) \) for the embedding function and subsequently derive the strip width:
\begin{equation}\label{sect 2: Ln}
	L_{\text{I}}(T, \sigma)=2\int_0^{z_{\text{max}}}S'(z) dz, \ \ \text{for}\ T< T_c, \ \sigma=1, 
\end{equation}
where \( z_{\text{max}} \) is the turning point that satisfies \( S'(z_{\text{max}})=\infty\). This point is illustrated by the red point in Figure \ref{GravityDual}. Typically, \( L_{n}(T, \sigma) \) is not well-defined for all ranges of tension; its domain determines the critical tension \( T_c \). We will see an exact example in Section 3.1.

It is important to note that the extrinsic curvature \( K_{ij} \) changes sign when crossing the EOW brane. Consequently, the tension \( T \) and the pre-factor of the stress tensor \( \sigma \) in the NBC (\ref{sect 2: NBC}) will change sign as well. As a result, the gravity dual for \( T > T_c, \sigma = 1 \) is the complement of the gravity dual for \( T < T_c, \sigma = -1 \). Therefore, we can express the width \( L \) for brane tension greater than the critical value as:
\begin{equation}\label{sect 2: Lp}
	L_{\text{II}}(T,\sigma)=\beta-L_\text{I}(-T, -\sigma) , \ \ \text{for}\ T> T_c, \ \sigma=1,
\end{equation}
where \( \beta \) is the bulk period defined earlier. \( L_{\text{II}}(T, \sigma) \) serves as the analytic continuation of \( L_{\text{I}}(T, \sigma) \) from \( T < T_c \) to \( T > T_c \), and it satisfies the following self-consistency limit:
\begin{equation}\label{sect 2: LnLp}
	L_{\text{II}}(T_c^+,1)=\beta-L_\text{I}(-T_c^-, -1)=L_\text{I}(T_c^-, 1),
\end{equation}
where $T_c^+=T_c+0^+$ and $T_c^-=T_c+0^-$. 
 
To derive the holographic Casimir amplitude \(\kappa_1\), we refer to equations (\ref{sect2: kappa1}, \ref{sect 2: Ln}, \ref{sect 2: Lp}). Additionally, to explore the holographic bound of \(-\kappa_1/C_D\) ( \ref{Intro: Casimir bound}), we require information about the displacement operator. The expressions for \(C_D\) in lower dimensions are given as follows:
\begin{align}\label{sect2: CD 2d 3d 4d}
C_D=\begin{cases}
  \frac{c}{2\pi^2},& \text{for} \ d=2,\\
 \frac{32}{\pi  \left(2 \tan ^{-1}\left(\tanh \left(\frac{\rho }{2}\right)\right)+\frac{\pi }{2}\right)},&
\text{for} \ d=3,\\
  \frac{120 e^{-\rho } \cosh (\rho )}{\pi ^2},&
\text{for} \ d=4,
\end{cases}
\end{align}
where $c=24 \pi$ is the central charge of 2D CFTs, and $\rho$ is related to the brane tension $T=(d-1)\tanh(\rho)$. Recall that we have set $16\pi G_N=1$. The case for general dimensions can be found in \cite{Miao:2024ddp}. One can check that $C_D$ (\ref{sect2: CD 2d 3d 4d}) decreases with the brane tension $T=(d-1)\tanh(\rho)$ for $d\ge 3$.

\subsection{Casimir bound from holographic g-theorem}

In this subsection, we explore an intriguing relationship between the holographic bound of the Casimir effect and the holographic g-theorem \cite{Takayanagi:2011zk,Fujita:2011fp}. By applying the null energy condition to the brane-localized matter fields, we derive the holographic g-theorem in AdS soliton and utilize it to demonstrate the holographic bound of the Casimir effect in two dimensions. However, in higher dimensions, the holographic g-theorem is not sufficient to establish the holographic bound. We will address this issue with specific matter fields in the next section.

We begin by examining the scenario where \(T \leq T_c\) and \(\sigma = 1\). For simplicity, we will focus on the left part of the EOW brane, noting that the analysis for the right part is similar. In the case of general brane-localized matter fields, the EOW brane may exhibit asymmetry, as discussed in Section 3.2. However, this asymmetry does not affect our analysis below. That is because we only require the general properties of the g-function, rather than the specific expressions for the embedding function of the EOW brane.

We choose the following null vector on the left part of the EOW  brane:
\begin{align}\label{sect 2: null vector}
l^{i}=\Big(\frac{\sqrt{z}}{\sqrt{f(z) S'(z)^2+\frac{1}{f(z)}}},\sqrt{z},0,...,0\Big).
\end{align}
By applying the null energy condition on the brane, we arrive at the following inequality:
\begin{align}\label{sect 2: NBC NEC}
\Big( K_{ij}-(K-T) h_{ij} \Big) l^il^j= T_{\text{matter}\ ij}l^il^j\ge 0, 
\end{align}
which leads to 
\begin{align}\label{sect 2: NBC NEC1}
\frac{\sqrt{f} \left(f' S' \left(f^2 \left(S'\right)^2+3\right)+2 f S''\right)}{2 \left(f^2 \left(S'\right)^2+1\right)^{3/2}} \ge 0.
\end{align}
Next, we define the \(g\)-function as follows:
\begin{align}\label{sect 2: g function}
g(z)=\frac{-f(z)^{3/2} S'(z)}{\sqrt{f(z)^2 S'(z)^2+1}}.
\end{align}
Using this definition, we can rewrite the inequality from above in the form of the \(g\)-theorem:
\begin{align}\label{sect 2: g theorem}
g'(z)\le 0.
\end{align}
This indicates that the \(g\)-function decreases under the boundary RG flow, as represented by the scale \(z\). At the AdS boundary, the \(g\)-function simplifies to the brane tension:
\begin{align}\label{sect 2: g function bdy}
g(0)=\tanh(\rho)=\frac{T}{d-1}. 
\end{align}
In the deepest bulk, where \(S'(z_{\text{max}}) = \infty\), the \(g\)-function becomes
\begin{align}\label{sect 2: g function bulk}
g(z_{\text{max}})=-\sqrt{f(z_{\text{max}})}=\frac{T_{\text{eff}}}{d-1},
\end{align}
where we have defined an effective brane tension $T_{\text{eff}}=(d-1)\tanh(\rho_{\text{eff}})$. The g-theorem (\ref{sect 2: g theorem}) implies 
\begin{align}\label{sect 2: g function tension}
T \ge T_{\text{eff}},
\end{align}
where the inequality is saturated in the absence of brane‑localized matter fields.

From (\ref{sect 2: Ln}) and (\ref{sect 2: g function}), we derive the left strip width for $T< T_c$
\begin{align}\label{sect 2: key trick 1}
L_{\text{I}\ L}&=\int_0^{z_{\text{max}}} dz S'(z) =\int_0^{z_{\text{max}}} dz \frac{-g(z)}{f(z) \sqrt{f(z)-g(z)^2}}.
\end{align}
Note that the integrand decreases with $g$:
\begin{align}\label{sect 2: key trick dg}
\frac{d}{dg(z)}\frac{-g(z)}{f(z) \sqrt{f(z)-g(z)^2}}=-\frac{1}{\left(f(z)-g(z)^2\right)^{3/2}}<0.
\end{align}
Recall that the g-theorem indicates that \(g(z_{\text{max}}) < g(z)\). Consequently, we can conclude:
\begin{align}\label{sect 2: key trick 2}
L_{\text{I} \ L}\le \int_0^{z_{\text{max}}} dz \frac{\sqrt{f(z_{\text{max}})}}{f(z) \sqrt{f(z)-f(z_{\text{max}})}}=\frac{L_0(T_{\text{eff}})}{2},
\end{align}
where \(L_0(T_{\text{eff}})\) represents the strip width with the effective tension \(T_{\text{eff}}\) in the absence of brane-localized matter fields. The results above also apply to the right part of the EOW brane. In total, we have:
\begin{align}\label{sect 2: key trick total L}
L_{\text{I}}(T<T_c, \sigma=1)=L_{\text{I} \ L}+L_{\text{I} \ R}\le L_0(T_{\text{eff}}).
\end{align}

Next, we consider the case where \(T > T_c\) and \(\sigma = 1\). Recall that \(L_\text{II}(T, \sigma = 1) = \beta - L_\text{I}(-T, \sigma = -1)\). For \(\sigma = -1\), the brane-localized matter fields become ghosts and violate the null energy conditions. Consequently, all of the previous inequalities change signs, leading to $L_\text{I}(-T, \sigma = -1) \ge L_0(-T_{\text{eff}})$, which yields:
\begin{align}\label{sect 2: key trick 3}
L_\text{II}(T>T_c, \sigma=1)=\beta-L_\text{I}(-T, \sigma=-1)\le \beta-L_0(-T_{\text{eff}})=L_0(T_{\text{eff}}).
\end{align}
By combining (\ref{sect 2: key trick total L}) and (\ref{sect 2: key trick 3}), we obtain the following inequality for general brane tensions: 
\begin{align}\label{sect 2: key trick 4}
L(T, \sigma=1) \le L_0(T_{\text{eff}}). 
\end{align} 

In the case of \(d = 2\), both \(L_0(T_{\text{eff}}) = \pi\) and \(C_D=c/(2\pi^2)\) are independent of the brane tension. Consequently, we have 
$\kappa_1|_{\sigma=1}=L^2\le L_0^2=\kappa_1|_{\sigma=0}$ and 
\begin{align}\label{sect 2: key trick 3d}
-\frac{\kappa_1}{C_D}|_{\sigma=1}\ge -\frac{\kappa_1}{C_D}|_{\sigma=0}\ge -\frac{\pi^3}{12} , \ \ \text{for} \ d=2.
\end{align}
This indicates that brane-localized matter fields typically reduce the Casimir effect and thus satisfy the holographic bound for 2D theories. Note that the original bound (\ref{Intro: Casimir bound}) focuses on BCFTs; here we extend it to theories without boundary conformal symmetries, as the brane-localized matter fields typically violate boundary conformal symmetries.  

The case in higher dimensions differs significantly from lower dimensions. We will first discuss the Casimir amplitude, denoted as \(\kappa_1 = L^d\). According to \cite{Miao:2024gcq, Miao:2025utb}, \(L_0(T_{\text{eff}})\) decreases with increasing brane tension for dimensions \(d \ge 3\). Recall that the holographic g-theorem indicates that \(T \ge T_{\text{eff}}\). Thus, we have
\begin{align}\label{sect 2: key trick 5}
L_0(T) \le L_0(T_{\text{eff}}). 
\end{align} 
Comparing this with (\ref{sect 2: key trick 4}), we cannot determine which of \(L(T)\) or \(L_0(T)\) is larger. In other words, the holographic g-theorem does not provide clarity on whether brane-localized matter fields reduce or increase the Casimir amplitude \(\kappa_1 = L^d\). We need additional information about the specific matter fields involved. Similarly, we remain uncertain about how the norm of the displacement operator \(C_D\) varies under the boundary RG flow. Generally, for non-CFTs in a half space, \(C_D\) is no longer constant:
\begin{align}\label{sect 2: CD nonCFT}
\langle T_{nn}(y) T_{nn}(0) \rangle= \frac{C_D( m y)}{|y|^{2d}},
\end{align} 
where $m$ represents the energy scale. Therefore, it is unclear how to generalize the holographic bound of $-\kappa_1/C_D$ to non-CFTs. However, if we adopt the effective central charge under the RG flow as \(C_D|_{\sigma=1, T} = C_D|_{\sigma=0, T_{\text{eff}}}\), we can extend the holographic bound to non-CFTs. By utilizing (\ref{sect 2: key trick 4}) and (\ref{Intro: Casimir bound}), we arrive at the conclusion:
\begin{align}\label{sect 2: new bound}
-\frac{\kappa_1}{C_D}|_{\sigma=1,T} \ge -\frac{\kappa_1}{C_D}|_{\sigma=0, T_{\text{eff}}}\ge \lim_{T\to T_{\text{min}}} \left(-\frac{\kappa_1}{C_D}\right)|_{\text{holo}}.
\end{align} 
Another possible choice is to take the effective central charge without matter fields, i.e., $C_D|_{\sigma=1, T}=C_D|_{\sigma=0, T}$. We will discuss this case in the following sections. However, we again stress that it remains an open question how to generalize the holographic bound for the Casimir effect to non-CFTs.

\section{Holographic Casimir effect of strips}

In this section, we study the Casimir effect of holographic strips with brane-localized scalars. To simplify our analysis, we consider the potential given by \( V =\frac{1}{2}m^2 \phi^2 \). Near the AdS boundary of the EOW brane, the brane-localized scalar exhibits the following asymptotic behavior:
\begin{align}\label{sect 3: scalar AdS bdy}
\phi=\lambda z^{d-1-\Delta} + \#\langle O \rangle z^{\Delta}+...
\end{align} 
where $\lambda$ is the boundary source, $\#$ denotes a constant factor, $O$ is the dual boundary operator, and $\Delta=\frac{d-1}{2}+\sqrt{(\frac{d-1}{2})^2+m^2 \cosh^2(\rho)}$ is the conformal dimension of $O$. By introducing scalar fields on the EOW brane, we effectively consider the following boundary deformations for the BCFT action:
\begin{align}\label{sect 3: action deformation}
I_{\text{def}}=I_{\text{BCFT}}+\int_{\partial M} d^{d-1}x \ \lambda O.
\end{align} 
The types of deformations are classified as follows:
\begin{align}\label{sect 3: deformation type}
& \text{relevant}, \ \ \ \ \text{for}\ m^2<0,\ \Delta<d-1\nonumber\\
& \text{marginal}, \ \ \ \text{for}\ m^2=0, \ \Delta=d-1\\
& \text{irrelevant}, \ \ \text{for}\ m^2>0, \ \Delta>d-1. \nonumber
\end{align}
For \( m^2 > 0 \), the scalar (\ref{sect 3: scalar AdS bdy}) generally diverges at the AdS boundary. To address this divergence, we focus on the case where the boundary source is zero, i.e., \( \lambda = 0 \). However, as we will establish in Section 3.2, there are no scalar solutions for \( m^2 > 0 \) when \( \lambda = 0 \). Therefore, we will concentrate on relevant and marginal deformations with $m^2\le 0$ in this section. Additionally, we require that \( m^2 \) satisfies the Breitenlohner–Freedman (BF) bound:
\begin{align}\label{sect 3: BF bound}
m^2\ge -(\frac{d-1}{2})^2\text{sech}^2(\rho). 
\end{align} 

Relevant deformations typically drive a theory away from its UV conformal fixed point under RG flow, generally breaking conformal invariance along the flow, although the theory may approach another CFT in the infrared. In contrast, exactly marginal deformations preserve conformal invariance and locally parametrize a continuous family of CFTs. For this new family of CFT where \( m^2 = 0 \), the norm of the displacement operator \( C_D \) and the ratio \( -\kappa_1/C_D \) are well-defined. In the next subsection, we will demonstrate that this new family of CFT satisfies the holographic bound of the Casimir effect stated in (\ref{Intro: Casimir bound}). However, for the cases of relevant deformations, \( C_D \) is not well-defined. Therefore, we will concentrate solely on the Casimir amplitude \( \kappa_1 \). Our findings indicate that \( \kappa_1 \) can increase under holographic relevant deformations.

\subsection{Massless scalar}

We first consider the massless free brane-localized scalar field, for which a conserved quantity exists that simplifies the equations of motion. It corresponds to exactly marginal deformations, leading to a new family of BCFT. Because the scalar field takes different values at the strip's two boundaries, it can effectively model the mixed boundary conditions of BCFTs.

Let us first examine the case where $T\le T_c, \sigma=1$. We denote the brane-localized scalar as \(\phi(z)\). The scalar EOM (\ref{sect 2: matter EOM}) with $V(\phi)=0$
\begin{equation}\label{sect 2: scalar EOM}
\partial_z \Big(\sqrt{-h} h^{zz} \phi'(z)\Big)=0, 
\end{equation}
yields a conserved quantity
\begin{equation}\label{sect 2: scalar conserved quantity}
q=\sqrt{-h} h^{zz} \phi'(z)=\frac{ \phi'(z)}{ z^{d-2} \sqrt{\frac{1}{f(z)}+f(z) S'(z)^2}}.
\end{equation}
For the left part of EOW brane, the $zz$-component of the NBC (\ref{sect 2: NBC}) provides:
\begin{equation}\label{sect 2: NBC zz}
T+(d-1) \frac{f^{\frac{3}{2}}S'(z)}{\sqrt{1+f^2 S'(z)^2}}=\frac{\sigma}{2}\ \frac{z^2 f(z)\phi'(z)^2}{1+f(z)^2 S'(z)^2}.
\end{equation}
From these two equations, we can solve for \(T \le T_c\):
\begin{equation}\label{sect 2: dphi}
\phi '(z)=\frac{2 (d-1) q z^{d-2}}{\sqrt{4 (d-1)^2 f(z)-\left(q^2 \sigma  z^{2 d-2}-2 T\right)^2}},
\end{equation}
and 
\begin{equation}\label{sect 2: dS}
S'(z)=\frac{q^2 \sigma  z^{2 d-2}-2 T}{f(z) \sqrt{4 (d-1)^2 f(z)-\left(q^2 \sigma  z^{2 d-2}-2 T\right)^2}}.
\end{equation}
It can be verified that the above two equations also satisfy the other components of the NBC (\ref{sect 2: NBC}).
The turning point is derived from the condition \(S'(z_{\text{max}}) = \infty\), leading to:
\begin{equation}\label{sect 2: zmax}
2 (d-1)\sqrt{ f(z_{\text{max}})}=q^2 \sigma  z_{\text{max}}^{2 d-2}-2 T=-2T_{\text{eff}}.
\end{equation}
Here, \(T_{\text{eff}}\) represents the effective brane tension. For \(\sigma = 1\), the above equation holds only if 
\begin{align}\label{sect 2: Tc 1}
T=-(d-1)\sqrt{1-z_{\text{max}}^d}+\frac{1}{2}q^2  z_{\text{max}}^{2 d-2}\le \frac{q^2}{2}, 
\end{align}
which suggests a critical tension
\begin{align}\label{sect 2: Tc massless scalar}
T_c =\frac{q^2}{2}. 
\end{align}
According to Sect. 2.1, the strip width is defined as
\begin{equation}
L(T)=
\begin{cases}
L_\text{I}(T, 1)=2\int_0^{z_{\text{max}}}S'(z)dz, & T< T_c=q^2/2,\\
L_\text{II}(T, 1)=\beta-L_\text{I}(-T, -1), & T> T_c=q^2/2.
\end{cases}
\end{equation}
Note that for \(q^2 \ge 2(d-1)\), we have \(T = (d-1) \tanh(\rho) \le T_c\) and \(L\) is always given by \(L_\text{I}\).

By substituting (\ref{sect 2: zmax}) into (\ref{sect 2: dS}), we obtain for $T<T_c, \sigma=1$:
\begin{align}\label{sect 2: dS1}
S'(z)&=\frac{\sqrt{f(z_{\text{max}}) }-g(z)+g(z_{\text{max}}) }{f(z) \sqrt{ f(z)- f(z_{\text{max}})+g(z_{\text{max}})^2-g(z)^2 }}\nonumber\\
&\le \frac{\sqrt{ f(z_{\text{max}}) } }{f(z) \sqrt{ f(z)- f(z_{\text{max}}) } },
\end{align}
where $g(z)$ is the g-function defined in (\ref{sect 2: g function}):
\begin{align}\label{sect 3: g function massless}
g(z)=\frac{2 T-q^2 \sigma  z^{2 d-2}}{2(d-1)}.
\end{align}
We can verify directly that it satisfies the holographic g-theorem $g'(z)\le 0$ for a healthy scalar with $\sigma=1$. From (\ref{sect 2: dS1}) and  \(L = 2\int_0^{z_{\text{max}}} S'(z)\) for $T< T_c$, we obtain 
\begin{align}\label{sect 3: width inequality 1}
L(T) \le 2\int_0^{z_{\text{max}}} dz \frac{\sqrt{f(z_{\text{max}})}}{f(z) \sqrt{f(z)-f(z_{\text{max}})}}=L_0(T_{\text{eff}}),
\end{align}
where \( L \) and \( L_0 \) denote the strip widths with and without brane-localized scalar fields, respectively. As discussed in Section 2.2, the inequality \( L(T) \leq L_0(T_{\text{eff}}) \) also holds for \( T \geq T_c \), and we will not repeat the argument here for simplicity.

\begin{figure}[htbp]
  \centering
\includegraphics[width=0.52\textwidth]{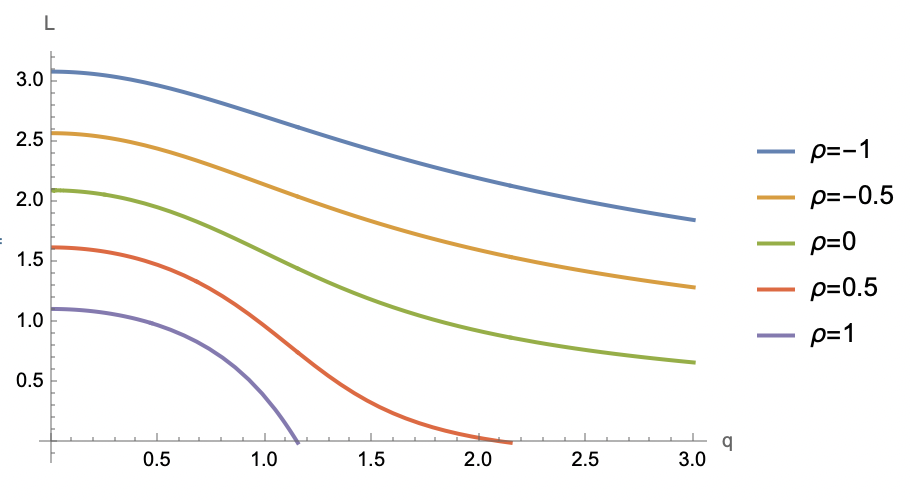} \includegraphics[width=0.45\textwidth]{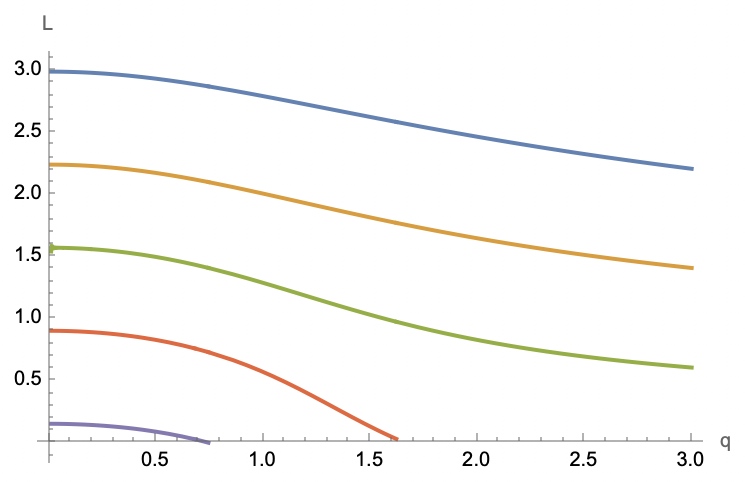}
 \caption{The strip width \( L \) decreases as the scalar parameter \( q \) increases. The left and right figures are for $d=3,4$, respectively. The condition \( L \ge 0 \) may impose an upper limit on \( q \) for a fixed \( \rho \).} 
 \label{L3d4d}
\end{figure}

As discussed in Section 2.2, the holographic g-theorem indicates that \( L(T) \leq L_0(T_{\text{eff}}) \). However, it does not specify which is greater between \( L(T) \) and \( L_0(T) \). For a brane-localized free scalar field, we can analytically demonstrate the following inequality for negative tensions and numerically validate it for general tensions: \begin{align}\label{sect 3: width inequality 2}
 L(T) \leq L_0(T), 
\end{align} 
This inequality implies that brane-localized free scalar fields decrease the Casimir amplitude \( \kappa_1 = L^d \) for a fixed brane tension \( T \). The proof of inequality (\ref{sect 3: width inequality 2}) for negative brane tension can be found in Appendix A. We also conducted extensive numerical verifications of this inequality for various brane tensions. Refer to Fig. \ref{L3d4d} for dimensions \( d = 3, 4 \), which illustrates that the strip width \( L \) decreases as the scalar parameter \( q \) increases. Moreover, the condition \( L \ge 0 \) may impose an upper limit on \( q \) for a fixed value of \( \rho \).

\begin{figure}[htbp]
  \centering
\includegraphics[width=0.52\textwidth]{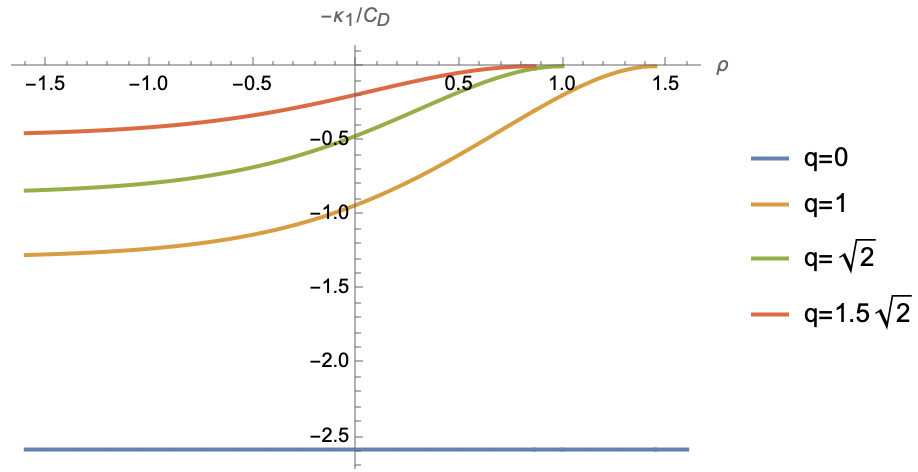} \includegraphics[width=0.42\textwidth]{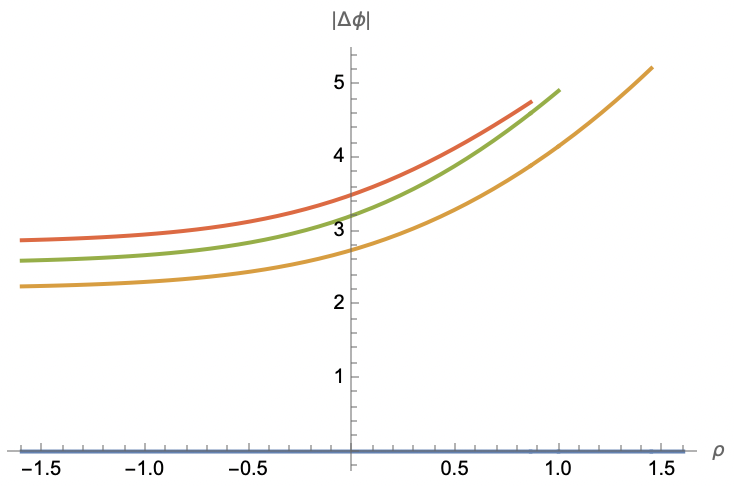}
 \caption{Casimir ratio $-\kappa_1/C_D$ (Left) and scalar difference $|\Delta \phi|$ (Right) for 2D strip. demonstrates that the scalar difference $|\Delta \phi|$ increases as the parameter $q$ increases. The blue, orange, green and red curves correspond to $q=0,1, \sqrt{2}, 1.5 \sqrt{2}$, respectively.  Conversely, the left figure indicates that the ratio $-\kappa_1/C_D$ increases with increasing $q$ and thus the scalar difference $|\Delta \phi|$. Overall, these two figures suggest that the brane-localized scalar diminishes the Casimir effect and adheres to the holographic lower bound (\ref{Intro: Casimir bound}), represented by the blue line of the left figure.} 
 \label{stripCasimir2d}
\end{figure}

 \begin{figure}[htbp]
  \centering
\includegraphics[width=0.52\textwidth]{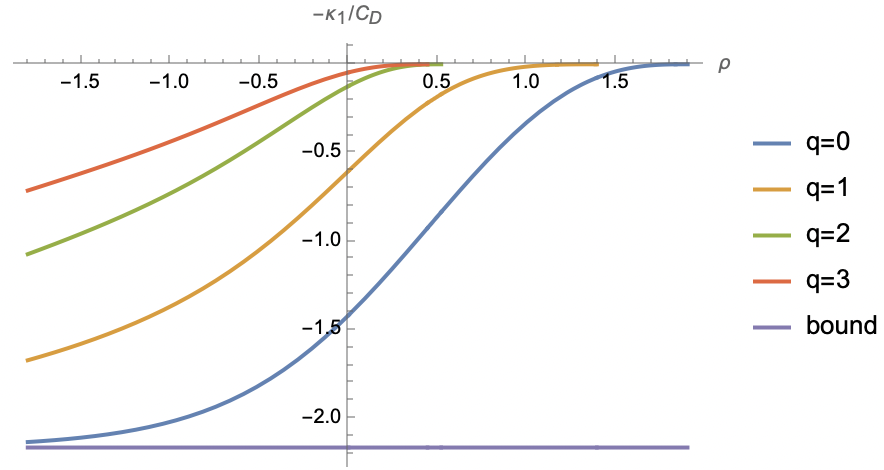} \includegraphics[width=0.42\textwidth]{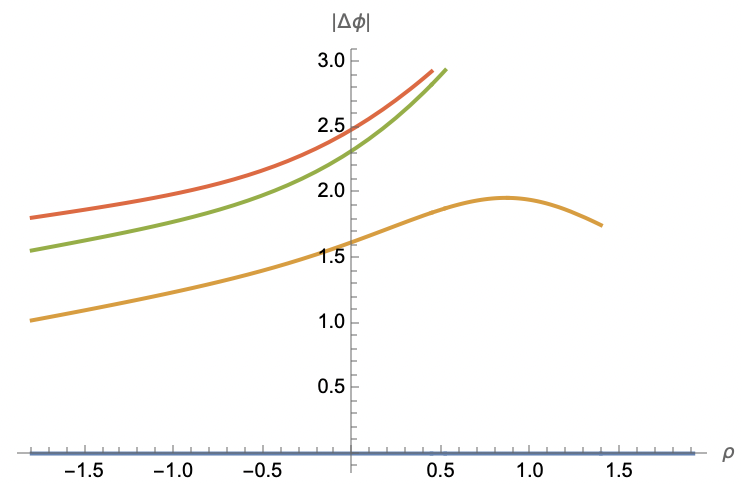}
 \caption{Casimir ratio $-\kappa_1/C_D$ (Left) and scalar difference $|\Delta \phi|$ (Right) for 3D strip. The right figure demonstrates that the scalar difference $|\Delta \phi|$ increases as the parameter $q$ increases. The blue, orange, green and red curves correspond to $q=0,1,2,3$, respectively.  Conversely, the left figure indicates that the ratio $-\kappa_1/C_D$ increases with increasing $q$ and thus the scalar difference $|\Delta \phi|$. Overall, these two figures suggest that the brane-localized scalar diminishes the Casimir effect and adheres to the holographic lower bound (\ref{Intro: Casimir bound}), represented by the purple line of the left figure.} 
 \label{stripCasimir3d}
\end{figure}

 \begin{figure}[htbp]
  \centering
\includegraphics[width=0.52\textwidth]{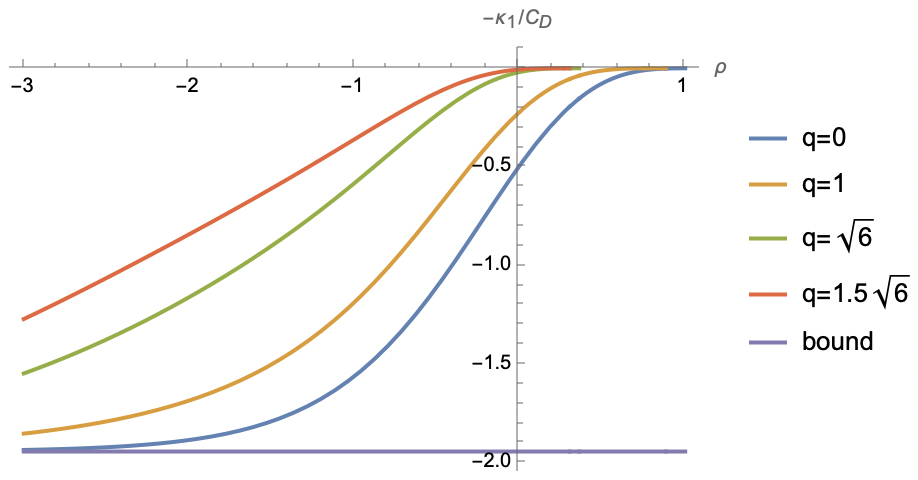} \includegraphics[width=0.42\textwidth]{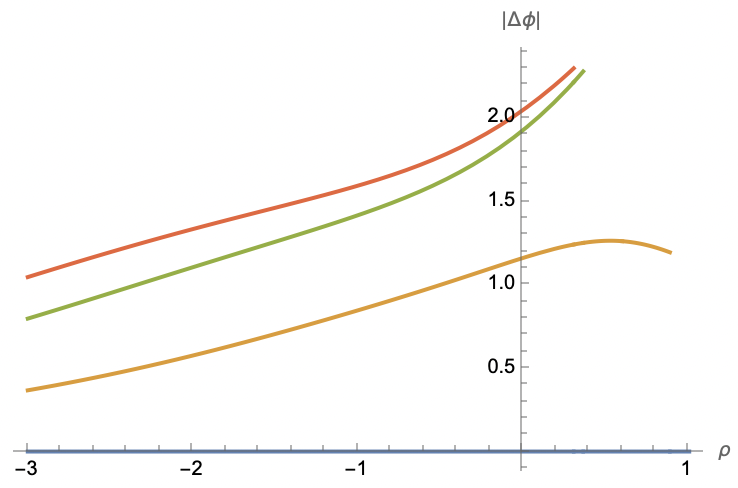}
 \caption{Casimir ratio $-\kappa_1/C_D$ (Left) and scalar difference $|\Delta \phi|$ (Right) for 4D strip. The right figure demonstrates that the scalar difference $|\Delta \phi|$ increases as the parameter $q$ increases. The blue, orange, green and red curves correspond to $q=0,1,\sqrt{6}, 1.5\sqrt{6}$, respectively.  Conversely, the left figure indicates that the ratio $-\kappa_1/C_D$ increases with increasing $q$ and thus the scalar difference $|\Delta \phi|$. Overall, these two figures suggest that the brane-localized scalar diminishes the Casimir effect and adheres to the holographic lower bound (\ref{Intro: Casimir bound}), represented by the purple line of the left figure.} 
 \label{stripCasimir4d}
\end{figure}

For a massless brane-localized scalar, the boundary deformation is marginal, leading to a new family of BCFTs. Therefore, the norm of the displacement operator \( C_D \) is well-defined. To derive \( C_D \), it is more straightforward to consider a half-space rather than a strip. For a holographic half-space, the constant solution \( \phi = \lambda = \text{constant} \) satisfies the EOM \( \Box \phi = 0 \) and results in zero matter stress tensors. Consequently, the marginal deformation induced by this solution does not alter \( C_D \). In other words, \( C_D \) for a massless brane-localized scalar remains given by  (\ref{sect2: CD 2d 3d 4d}). Combining this information with \( \kappa_1 = L^d \) and inequality (\ref{sect 3: width inequality 2}) leads to 
\begin{align}\label{sect 2: ratio larger} 
-\frac{\kappa_1}{C_D} \bigg|_{\sigma=1} \geq -\frac{\kappa_1}{C_D} \bigg|_{\sigma=0}, 
\end{align} 
which indicates that the new BCFTs induced by a massless brane-localized scalar still satisfy the holographic bound for the Casimir effect. We will verify this result numerically below.

We present the ratio \(-\kappa_1/C_D\) and the scalar difference \(|\Delta \phi|\) in Figures \ref{stripCasimir2d}, \ref{stripCasimir3d}, and \ref{stripCasimir4d} for dimensions \(d = 2, 3, 4\), respectively. Note that \(L \ge 0\) establishes an upper limit on the brane tension. This is why the curves for the ratios terminate at sufficiently large values of the tension parameter \(\rho\). These figures demonstrate that the scalar difference \(|\Delta \phi|\), which corresponds to the mixed boundary conditions, increases with \(q\). Additionally, the ratio \(-\kappa_1/C_D\) increases as both \(q\) and the scalar difference \(|\Delta \phi|\) increase. Consequently, massless brane-localized scalar fields generally reduce the Casimir effect and adhere to the holographic bound for the Casimir effect for BCFTs.

\subsection{Massive scalar}

We will now examine the holographic Casimir effect for massive brane-localized scalar fields. We will prove a no-hair theorem for cases where the mass squared is positive, specifically \( m^2 > 0 \), and with the boundary source set to \( \lambda = 0 \). Additionally, we will numerically verify that a massive brane-localized scalar field with \( m^2 < 0 \) can enhance the Casimir amplitude \( \kappa_1 \).

Unlike the massless scalar case, a massive scalar field has no conserved quantity, so we must use a numerical approach. For the sake of convenience in numerical calculations, we define the embedding function of the EOW brane as follows:
\begin{align}\label{sect 3.2: massive scalar brane}
\text{brane}:\ z=Z(x).
\end{align}
To simplify the analysis, we will focus on negative brane tension, \( T \leq 0 \), and consider a ghost-free scalar with \( \sigma = 1 \). We will also set \( z_h = 1 \) and assume that \( \phi = \phi(x) \). By solving the NBC (\ref{sect 2: NBC}) and scalar EOM (\ref{sect 2: matter EOM}) with $V=\frac{1}{2}m^2 \phi^2$, we obtain two independent equations:
\begin{align} 
(d-1)
\left[
\tanh\rho
+
\frac{f(Z)^{3/2}}
{\sqrt{f(Z)^2+\left(Z'\right)^2}}
\right]
={}&
\frac{\sigma}{2}
\left[
\frac{Z^2 f(Z)}
{f(Z)^2+\left(Z'\right)^2}
\left(\phi'\right)^2
-
m^2\phi^2
\right],
\label{sect 3.2: EOM1}
\\[2mm]
\frac{d}{dx}\left[\frac{Z^{2-d}\sqrt{f(Z)}}{\sqrt{f(Z)^2+\left(Z'\right)^2}}\phi'\right]
={}&
\frac{m^2\sqrt{f(Z)^2+\left(Z'\right)^2}}{Z^d\sqrt{f(Z)}}\phi .
\label{sect 3.2: EOM2}
\end{align}

Expanding around the turning point of the EOW brane, we have:
\begin{align}\label{sect 3.2: solution around turning point} 
Z(x)=z_{\max}+ O(x^2),\ \  \phi(x)=\phi_0+ \phi_1 x+O(x^2),
\end{align}  
where we have used \(Z'(0) = 0\) at the turning point \(z = z_{\max}, x = 0\). Solving the EOMs (\ref{sect 3.2: EOM1},\ref{sect 3.2: EOM2}) around the turning point, we derive
\begin{align} \label{sect 3.2: relation}
T+(d-1)\sqrt{f_m}
-\frac{z_{\max}^2}{2f_m}\phi_1^2
+\frac{1}{2}m^2\phi_0^2=0,
\end{align}
where $f_m\equiv f(z_{\max})=1-z_{\max}^d$. For \(\phi_1\) to be real, we require:
\begin{align} \label{sect 3.2: constraint phi1}
T+(d-1)\sqrt{f_m}+\frac{1}{2}m^2\phi_0^2\ge 0,
\end{align}
which implies that, for \(m^2 < 0\), \(\phi_0^2\) cannot be too large. By solving for $Z(x)$ to higher orders, we find that 
\begin{align} \label{sect 3.2: Z higher orders} 
Z(x)-Z(-x)=-\frac{2m^2 f_m^{3/2}}{3z_{\max}}\,\phi_0\phi_1 x^3+O(x^5). 
\end{align} 
This indicates that, in contrast to the massless case, $Z(x)$ lacks symmetry for massive scalars generally.
 
The case \( m^2 > 0 \) is special and requires a detailed discussion. For \( m^2 > 0 \), the scalar field (\ref{sect 3: scalar AdS bdy}) generally diverges at the AdS boundary. To eliminate this divergence, we set the source to zero, i.e., \( \lambda = 0 \). Under this condition, we can demonstrate a no-hair theorem, which states that the scalar field cannot support a connected EOW brane linking the two strip boundaries. We rewrite the scalar EOM (\ref{sect 3.2: EOM2}) as
\begin{align} \label{sect 3.2: no hair 1} 
 -(U \phi')'+m^2 Y \phi=0
\end{align} 
where 
\begin{align} \label{sect 3.2: no hair U V} 
U=\frac{Z^{2-d}\sqrt{f(Z)}}{\sqrt{f(Z)^2+\left(Z'\right)^2}}>0, \  Y=\frac{\sqrt{f(Z)^2+\left(Z'\right)^2}}{Z^d\sqrt{f(Z)}}>0.
\end{align} 
Multiplying the equation by \( \phi \) and integrating by parts leads to the following result:
\begin{align} \label{sect 3.2: no hair 2} 
\int_{x_-}^{x^+} dx \Big( U \phi'^2+ m^2 Y \phi^2\Big)=U \phi \phi'\Big|^{x^+}_{x_-},
\end{align} 
where \( x_+ \) and \( x_- \) are the endpoints of the strip. The left-hand side of this equation is non-negative for \( m^2 > 0 \), while the right-hand side vanishes. By using (\ref{sect 3: scalar AdS bdy}) with \( \lambda = 0 \), along with conditions \( x \sim -\sinh(\rho) z \) and \( 2\Delta - (d-1) > 0 \) for \( m^2 > 0 \), we obtain:
\begin{align} \label{sect 3.2: no hair 3} 
U \phi \phi'\Big|^{x^+}_{x_-}\sim \lim_{z\to 0}  z^{2\Delta-(d-1)}\to 0.
\end{align} 
As a result, to satisfy (\ref{sect 3.2: no hair 2}), the scalar must vanish for $m^2>0$. Note that this proof relies solely on the scalar EOM and the asymptotic behavior of the scalar field, which means it applies to general tensions \( T \) and \( \sigma \). 

\begin{table}[htbp]
    \centering
    \begin{tabular}{c|ccc}
        \hline
        $m^2$
        & $z_{\max}^{\mathrm{opt}}$
        & $\phi_0^{\mathrm{opt}}$
        & $L_{\max}$ \\
        \hline
    -0.40 & 0.7489 & 0.0000 & 3.0789 \\
-0.30 & 0.7430 & 0.2921 & 3.0792 \\
-0.20 & 0.6968 & 1.0180 & 3.0986 \\
-0.10 & 0.6262 & 2.0686 & 3.1873 \\
0.00 & 0.7489 & 0.0000 & 3.0789 \\
        \hline
    \end{tabular}
    \caption{
        Maximum strip width obtained by optimizing over
        $z_{\max}$ and $\phi_0$ for
        $d=3$, $\rho=-1$ and $z_h=1$. 
        For $m^2=0$, the constant part of $\phi_0$ does not affect
        the geometry, and we choose the representative value $\phi_0=0$.
    }
    \label{tab:max-strip-width-d3}
\end{table}

Let us go on to discuss the case $m^2<0$. For the given theory parameters \( T < 0 \), \( m^2<0 \), and initial values \( z_{\max} \) and \( \phi_0 \), we can numerically solve  EOMs (\ref{sect 3.2: EOM1},\ref{sect 3.2: EOM2}) with the initial condition (\ref{sect 3.2: solution around turning point}). We can determine the strip width \( L \) using the formula \begin{align} \label{sect 3.2: to fix L} 
L = x_+ - x_-, \quad Z(x_+) = Z(x_-) = 0, 
\end{align} 
where \( x_+ > 0 \) and \( x_- < 0 \) are the endpoints of the strip. We aim to ascertain whether the presence of a massive brane-localized scalar can enhance the Casimir amplitude \( \kappa_1 = L^d \) compared to the case without scalars. To achieve this, we will adjust the initial values \( z_{\max} \) and \( \phi_0 \) within the constraint outlined in (\ref{sect 3.2: constraint phi1}) to maximize \( L \). As shown in Table \ref{tab:max-strip-width-d3} for \( d=3 \) and Table \ref{tab:max-strip-width-d4} for \( d=4 \), the largest strip width \( L_{\max} \) for \( m^2 < 0 \) may be greater than that for \( m^2 = 0 \). Thus, a negative \( m^2 \) can indeed enhance the Casimir amplitude \( \kappa_1 = L^d \). Note that we retain more decimal places in our numerical computations; for simplicity, only four decimal places are listed in the tables.

As discussed in Section 2, the dual theory is a non-CFT when \( m^2 \neq 0 \), and \( C_D \) is not well-defined in this scenario. If we set \( C_D(\sigma=1, m^2 \neq 0) \) equal to \( C_D(\sigma=0, T_{\text{eff}}) \), the holographic bound (\ref{Intro: Casimir bound}) still holds. However, if we define \( C_D(\sigma=1, m^2 \neq 0) \) as \( C_D(\sigma=0, T) \), the data in Table \ref{tab:max-strip-width-d3} and Table \ref{tab:max-strip-width-d4} suggest that the holographic bound (\ref{Intro: Casimir bound}) may be violated. For instance, if we choose \( \rho = -1 \) and \( m^2 = -0.1 \) for \( d=3 \), we find \( L_{\max} \approx  3.1873 \) and \( C_D(\sigma=0, T) \approx 14.4476 \), leading to a violation of the holographic bound (\ref{Intro: Casimir bound}):
\begin{align} \label{sect 3.2: violate bound 3d}
-\frac{\kappa_1}{C_D}|_{m^2=-0.10}\approx -2.24 < -\frac{\kappa_1}{C_D}|_{\text{lower bound}}\approx -2.17, \ \ \ \text{for}\ d=3.
\end{align}
Similarly, if we take \( \rho = -1 \) and \( m^2 = -0.10 \) for \( d=4 \), we get: 
\begin{align} \label{sect 3.2: violate bound 4d}
-\frac{\kappa_1}{C_D}|_{m^2=-0.10}\approx -2.74 < -\frac{\kappa_1}{C_D}|_{\text{lower bound}}\approx -1.94, \ \ \ \text{for}\ d=4.
\end{align}
It is important to emphasize that the initial bound (\ref{Intro: Casimir bound}) was proposed for BCFTs. Since \( C_D \) is not well-defined for non-BCFTs, it remains an open question how to generalize the bound (\ref{Intro: Casimir bound}) to encompass non-BCFTs.

\begin{table}[htbp]
    \centering
    \begin{tabular}{c|cccc}
        \hline
        $m^2$
        & $z_{\max}^{\mathrm{opt}}$
        & $\phi_0^{\mathrm{opt}}$
        & $L_{\max}$ \\
        \hline
 -0.40 & 0.7300 & 1.1265 & 3.0557 \\
-0.30 & 0.7040 & 1.4627 & 3.1122 \\
-0.20 & 0.6695 & 1.9922 & 3.2146 \\
-0.10 & 0.6154 & 3.1361 & 3.4384 \\
0.00 & 0.8050 & 0.0000 & 2.9889 \\
        \hline
    \end{tabular}
    \caption{
        Maximum strip width obtained by
        optimizing over $z_{\max}$ and $\phi_0$ for
        $d=4$, $\rho=-1$, and $z_h=1$.
        For $m^2=0$, the maximum occurs at the pure-tension endpoint
        $\phi_1=0$. The constant part of $\phi_0$ does not affect the
        geometry, and the representative value $\phi_0=0$ is chosen.
    }
    \label{tab:max-strip-width-d4}
\end{table}

To summarize, we find that the BCFTs dual to massless brane-localized scalar fields still obey the holographic bound (\ref{Intro: Casimir bound}) of the Casimir effect. In contrast, for non-BCFTs that are dual to massive brane-localized scalar fields, the value of \(C_D\) is not constant, and we concentrate on the Casimir amplitude \( \kappa_1 = L^d \). A negative squared mass can increase the Casimir amplitude for dimensions \(d \geq 3\), while it decreases the amplitude for \(d = 2\), as shown in Section 2.2. Conversely, a positive squared mass cannot support a connected EOW brane, and we demonstrate a no-hair theorem for this case when the scalar source is zero.

\section{Holographic Casimir effect of wedges}

This section investigates the holographic Casimir effect of wedges in AdS/BCFT with brane-localized scalar fields.

According to \cite{Miao:2024ddp}, the renormalized stress tensor for the BCFT in a wedge has the following form:
\begin{align}\label{sect 4: wedge Tij}
\langle T^{i}_{\ j} \rangle_{\text{wedge}}=\frac{f(\Omega)}{r^d} \text{diag}\Big(-(d-1),1,\dots,1 \Big),
\end{align}
where \(\Omega \) is the opening angle of the wedge, $f(\Omega)$ is the Casimir amplitude, and $r$ is the distance to the corner of the wedge. In the small-angle limit, where \(\Omega\) approaches 0, and in the large-radius limit, where \(r \Omega=L\) remains constant, a wedge configuration approaches that of a strip. To get the correct strip stress tensor in these limits, we derive
\begin{align}\label{sect 4: small angle} 
\lim_{\Omega \to 0} f(\Omega) \Omega^d =\kappa_1. 
\end{align}
It has been conjectured that AdS/BCFT with minimal brane tension establishes a lower bound for the wedge Casimir effect \cite{Miao:2024gcq, Miao:2025utb}:
\begin{align}\label{sect 4: Casimir bound} 
-\frac{f(\Omega)}{C_D} \ge \lim_{T\to T_{\text{min}}} \left(-\frac{f(\Omega)}{C_D}\right)|_{\text{holo}}, \ \text{for} \ 0<\Omega\le \pi.
\end{align} 
This section tests the proposal above by studying brane-localized scalar fields.

The wedge space is dual to a portion of the AdS soliton featuring a hyperbolic transverse space \cite{Miao:2024ddp}. The metric of the AdS soliton and the embedding function of the EOW brane are given by \cite{Miao:2024ddp}:
\begin{align}\label{sect 4: soliton}
&\text{metric}:\ ds^2=\frac{\frac{dz^2}{h(z)}+h(z) d\theta^2+\frac{dr^2+\sum_{\hat{i},\hat{j}=1}^{d-2} \eta_{\hat{i}\hat{j}} dy^{\hat{i}} dy^{\hat{j}}}{r^2}}{z^2}, \\
&\text{brane}:\ \theta=\theta(z), \label{sect 4: brane}
\end{align}
where $h(z)=1-z^2-c_1 z^d$. It produces the expected Casimir effect (\ref{sect 4: wedge Tij}), provided that \cite{Miao:2024ddp}: 
\begin{eqnarray}\label{sect 4: c1}
c_1=f(\Omega).
\end{eqnarray}
The angle period in bulk is given by
\begin{eqnarray}\label{sect 4: beta}
\beta=\frac{4\pi}{|h'(z_h)|}=\frac{4\pi z_h}{d+(2-d) z_h^2},
\end{eqnarray}
with $h(z_h)=0$ and $c_1=(1-z_h^2)/z_h^d$. On the AdS boundary, we have $0\le \theta\le \Omega$. From the NBC (\ref{sect 2: NBC}), we can determine the embedding function of the EOW brane and then derive the opening angle of the wedge as
\begin{eqnarray}\label{sect 4: opening angle 1} 
\Omega_{\text{I}}(T, \sigma)=2\int_0^{z_{\text{max}}} \theta'(z) dz, \ \text{for } T< T_c,
\end{eqnarray}
and 
\begin{eqnarray}\label{sect 4: opening angle 2} 
\Omega_{\text{II}}(T, \sigma)=\beta-\Omega_{\text{I}}(T\to -T, \sigma\to -\sigma), \ \text{for } T> T_c.
\end{eqnarray}

\subsection{Massless scalar}

Let us first study the massless brane-localized scalar. Similar to the case of strip, we derive a conserved quantity from the scalar EOM
\begin{align}\label{sect 4.1: q} 
q=\frac{\phi'(z)}{z^{d-2}\sqrt{\frac{1}{h(z)}+h(z)\theta'(z)^2}}.
\end{align} 
For the left part of EOW brane with $T\le T_c$, the $zz$-component of the NBC (\ref{sect 2: NBC}) yields:
\begin{equation}\label{sect 4.1: NBC zz}
T+(d-1) \frac{h(z)^{\frac{3}{2}}\theta'(z)}{\sqrt{1+h(z)^2 \theta'(z)^2}}=\frac{\sigma}{2}\ \frac{z^2 h(z)\phi'(z)^2}{1+h(z)^2 \theta'(z)^2}.
\end{equation}
From these two equations, we can solve for \(T \le T_c\):
\begin{equation}\label{sect 4.1: dphi}
\phi '(z)=\frac{2 (d-1) q z^{d-2}}{\sqrt{4 (d-1)^2 h(z)-\left(q^2 \sigma  z^{2 d-2}-2 T\right)^2}},
\end{equation}
and 
\begin{equation}\label{sect 4.1: dS}
\theta'(z)=\frac{q^2 \sigma  z^{2 d-2}-2 T}{h(z) \sqrt{4 (d-1)^2 h(z)-\left(q^2 \sigma  z^{2 d-2}-2 T\right)^2}}.
\end{equation}
By using \(\theta'(z_{\text{max}}) = \infty\), we derive the turning point
\begin{equation}\label{sect 4.1: zmax}
2 (d-1)\sqrt{ h(z_{\text{max}})}=q^2 \sigma  z_{\text{max}}^{2 d-2}-2 T
\end{equation}
For \(\sigma = 1\) and $c_1=(1-z_h^2)/z_h^d>0$, the above equation holds only if 
\begin{align}\label{sect 4.1: Tc 1}
T=-(d-1)\sqrt{1-z_{\max}^2-c_1 z_{\text{max}}^d}+\frac{1}{2}q^2  z_{\text{max}}^{2 d-2}\le \frac{q^2}{2} z_{h}^{2 d-2}, 
\end{align}
which yields the critical tension
\begin{align}\label{sect 4.1: Tc massless scalar}
T_c =\frac{q^2}{2}z_{h}^{2 d-2}.  
\end{align}
From the above equations, we can derive the opening angle of the wedge for $\sigma=1$ as
\begin{equation}\label{sect 4.1: angle}
\Omega(T,1)=
\begin{cases}
\Omega_{\text{I}}(T, 1)=2\int_0^{z_{\text{max}}}\theta'(z)dz, & T\le T_c,\\
\Omega_{\text{II}}(T, 1)=\beta-\Omega_{\text{I}}(-T, -1), & T\ge T_c.
\end{cases}
\end{equation}
Note that $z_{\max}$ and $S'(z)$ depend on $c_1=f(\Omega)$. Thus, the above equation is the inverse function of $f(\Omega)$. 

Let us first analyze the limit of minimal brane tension as \( T \to T_{\text{min}} = -(d-1) \), which establishes the lower bound of the ratio \( -f(\Omega)/C_D \) \cite{Miao:2025utb}. Following the approach outlined in \cite{Miao:2025utb}, we find that this ratio is independent of the scalar parameter \( q \) in the minimal tension limit. Consequently, the BCFTs dual to a massless brane-localized scalar still satisfy the holographic bounds of the Casimir effect. For \( d=2 \), a wedge can be obtained from a strip through conformal transformations. We will focus on the non-trivial cases where \( d \geq 3 \) below.

In the limit as \( x = \text{sech}^2(\rho) \to 0 \), we have
\begin{eqnarray}\label{sect 4.1: CD limit}
C_D=\frac{2 (d-1) \pi ^{\frac{1}{2}-\frac{d}{2}} \Gamma (d+2)}{d \Gamma \left(\frac{d+1}{2}\right)} x^{1-\frac{d}{2}}+O\left(x^{2-\frac{d}{2}}\right).
\end{eqnarray}
We denote \( f(\Omega) = \hat{f}(\Omega) x^{1 - \frac{d}{2}} \) and \( z_{\max} = z_0 \sqrt{x} \), which allows us to derive a finite ratio:
\begin{eqnarray}\label{sect 4.1: ratio limit}
\text{ra}(\Omega)=\lim_{x\to 0}( \frac{-f(\Omega)}{C_D})=\frac{-\hat{f}(\Omega)}{\frac{2 (d-1) \pi ^{\frac{1}{2}-\frac{d}{2}} \Gamma (d+2)}{d \Gamma \left(\frac{d+1}{2}\right)} }
\end{eqnarray} 
and a finite turning-point condition (\ref{sect 4.1: zmax}):
\begin{eqnarray}\label{sect 4.1: zmax limit}
\hat{f} z_0^d+z_0^2-1+q^2 O(x^{d-2})=0. 
\end{eqnarray} 
Note that the contributions from the scalar vanish in the minimal tension limit \( x \to 0 \) for \( d \geq 3 \).

By substituting \( c_1 = f = \hat{f} x^{1 - \frac{d}{2}} \) into equation (\ref{sect 4.1: angle}) and taking the limit as \( x \to 0 \), we arrive at the following equation: 
\begin{eqnarray}\label{sect 4.1: opening angle result limit} 
\Omega = \int_0^{1} dy \frac{2 z_0}{\sqrt{1 - y^2 z_0^2 + (z_0^2 - 1) y^d}} + q^2 O(x^{d-2}), 
\end{eqnarray} 
where the scalar terms vanish as \( x \to 0 \). From the previous expressions, we observe that \( z_0 \) is a function of the ratio \( \text{ra}(\Omega) = -f(\Omega)/C_D \). Consequently, the derived equation represents the inverse function of \( \text{ra}(\Omega) \) in the minimal tension limit. Importantly, in this limit, the ratio \( \text{ra}(\Omega) \) is independent of the massless brane-localized scalar. Thus, it is anticipated that the holographic BCFTs with massless brane-localized scalars will continue to adhere to the holographic bound of the Casimir effect for a wedge. This is numerically confirmed for cases with general tension where \( T > T_{\text{min}} \). 

\begin{figure}[htbp]
  \centering
\includegraphics[width=0.495\textwidth]{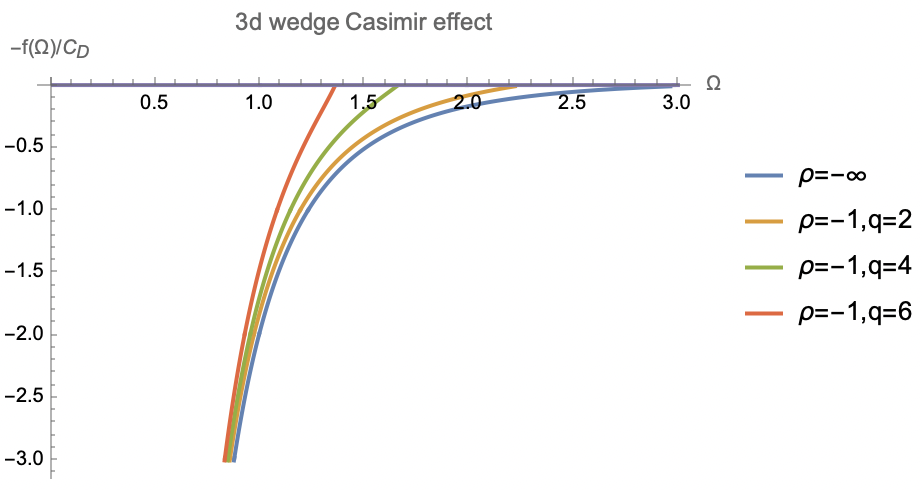} \includegraphics[width=0.495\textwidth]{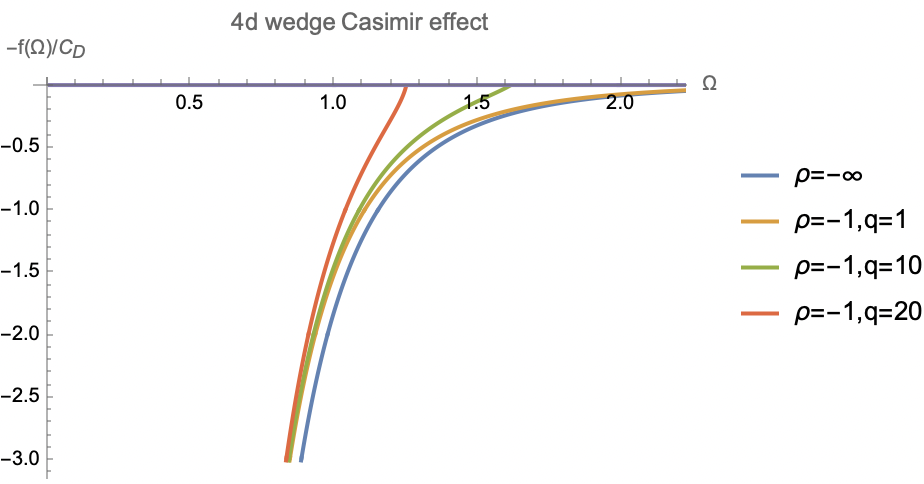}
 \caption{Casimir ratio $-f(\Omega)/C_D$ for 3d (Left) and 4d (Right) wedges for $\rho=-1$. They show the ratio $-f(\Omega)/C_D$ increases with the scalar parameter $q$, implying the massless brane-localized scalar diminishes the Casimir effect and adheres to the holographic lower bound (\ref{sect 4: Casimir bound}), represented by the blue curve of the figures.} 
 \label{wedgeCasimir}
\end{figure}

For instance, refer to Fig. \ref{wedgeCasimir} for $\rho=-1$, which illustrates that the ratio \( -f(\Omega)/C_D \) increases with the scalar parameter \( q \), thereby satisfying the holographic lower bound (\ref{sect 4: Casimir bound}). Notably, Fig. \ref{wedgeCasimir} also suggests that \( f(\pi) \neq 0 \) for \( q \neq 0 \), which is typical for mixed boundary conditions. Conversely, we have \( f(\pi) = 0 \) if we impose the same boundary conditions on both wedge boundaries. In this scenario, the wedge effectively becomes a half-space, which exhibits a vanishing Casimir effect for a BCFT.

\subsection{Massive scalar}

Let us examine the holographic wedges with massive brane-localized scalars. As in our strip analysis, a no-hair theorem applies to scalars with \( m^2 > 0 \) and \( \lambda = 0 \). Additionally, brane-localized scalars with negative values of \( m^2 \) can enhance the Casimir amplitude. Since the calculations are analogous to those for strips, we emphasize only the key points below.

For simplicity, we focus on the case where \( T < 0 \) and \( \sigma = 1 \). Assuming the embedding function of the EOW brane is given by \( z = Z(\theta) \), we obtain two independent equations from NBC and the scalar EOM:
\begin{equation}\label{sect 4.2: scalar EOM}
\frac{d}{d\theta}\left[\frac{Z^{2-d}\sqrt{h}}{\sqrt{h^2+\dot{Z}^2}}\dot{\phi}\right]
=\frac{m^2\sqrt{h^2+\dot{Z}^2}}{Z^d\sqrt{h}}\phi .
\end{equation}
and 
\begin{equation}\label{sect 4.2: NBC zz}
T+
\frac{(d-1)h^{3/2}}
     {\sqrt{h^2+\dot Z^2}}
=\frac{1}{2}
\left[
\frac{Z^2h}{h^2+\dot Z^2}\,\dot{\phi}^{\,2}
-m^2\phi^2
\right],
\end{equation}
where $\dot{f}=\partial_{\theta}f$, and $h=1-Z^2-c_1Z^d$. 

Expanding around the turning point of the EOW brane, we have:
\begin{align}\label{sect 4.2: solution around turning point} 
Z(\theta)=z_{\max}+ O(\theta^2),\ \  \phi(\theta)=\phi_0+ \phi_1 \theta+O(\theta^2),
\end{align}  
where we have used \(\dot{Z}(0) = 0\) at the turning point \(z = z_{\max}, \theta = 0\). Solving the EOMs (\ref{sect 4.2: scalar EOM},\ref{sect 4.2: NBC zz}) around the turning point, we derive a constraint equation for $\phi_1$:
\begin{equation}\label{sect 4.2: phi1}
\phi_1^2=\frac{h_m}{z_{\max}^2}\left[m^2\phi_0^2+2T+2(d-1)\sqrt{h_m}\right],
\end{equation}
where $h_m\equiv h(z_{\max})=1-z_{\max}^2-c_1z_{\max}^d$. The opening angle $\Omega$ of the wedge can be determined by 
\begin{align}\label{sect 4.2: fix Omega} 
\Omega=\theta_+-\theta_-,\   \ Z(\theta_+)=Z(\theta_-)=0.
\end{align}

Note that the scalar EOM (\ref{sect 4.2: scalar EOM}) takes the same form as that of a strip. Using the approach outlined in Section 3.2, we can directly derive that \(\phi = 0\) for \(m^2 > 0\) and \(\lambda = 0\). Therefore, we will concentrate on the non-trivial case when \(m^2 < 0\) in the following discussion.

\begin{table}[htbp]
    \centering
    \caption{The maximal Casimir amplitude $c_{1,\max}$ for
    $d=3$, $\Omega=\pi/2$, and $\rho=-1$.}
    \label{wedge Casimir 3d}
    \begin{tabular}{c|ccc}
        \hline\hline
        $m^2$ & $z_{\max}^{*}$ & $|\phi_0^{*}|$ & $c_{1,\max}$ \\
        \hline
 0.00 & 0.3645 & 0.0000 & 5.9285 \\
 -0.10 & 0.3203 & 1.6866 & 6.2071 \\
 -0.20 & 0.3618 & 0.3306 & 5.9298 \\
 -0.30 & 0.3645 & 0.0000 & 5.9285 \\
 -0.40 & 0.3645 & 0.0000 & 5.9285 \\
        \hline\hline
    \end{tabular}
\end{table}

\begin{table}[htbp]
    \centering
    \caption{The maximal Casimir amplitude $c_{1,\max}$ for
    $d=4$, $\Omega=\pi/2$, and $\rho=-1$.}
    \label{wedge Casimir 4d}
    \begin{tabular}{c|ccc}
        \hline\hline
        $m^2$ & $z_{\max}^{*}$ & $|\phi_0^{*}|$ & $c_{1,\max}$ \\
        \hline
  0.00 & 0.3899 & 0.0000 & 11.6011 \\
-0.10 & 0.2910 & 2.8068 & 16.4476 \\
-0.20 & 0.3338 & 1.6375 & 13.2491 \\
-0.30 & 0.3586 & 1.0685 & 12.1575 \\
-0.40 & 0.3756 & 0.6631 & 11.7284 \\
        \hline\hline
    \end{tabular}
\end{table}

Solving EOMs (\ref{sect 4.2: scalar EOM},\ref{sect 4.2: NBC zz}) with the initial conditions (\ref{sect 4.2: solution around turning point}, \ref{sect 4.2: phi1}), we can numerically determine \( Z(\theta) \) and the opening angle \( \Omega \) using equation (\ref{sect 4.2: fix Omega}).  For a given \( \Omega \), we adjust the parameters \( z_{\max} \) and \( \phi_0 \) to maximize the Casimir amplitude \( c_1 = f(\Omega) \). For instance, taking \( \Omega = \frac{\pi}{2} \), \( \rho = -1 \), and the range \( -\left(\frac{d-1}{2}\right)^2 \text{sech}^2(\rho) \le m^2 \le 0 \), we can reference Tables \ref{wedge Casimir 3d} and \ref{wedge Casimir 4d} to show the maximal Casimir amplitude \( c_{1,\max} \) for dimensions \( d = 3 \) and \( d = 4 \). These tables illustrate that, similar to the case of the strip, negative values of \( m^2 \) can enhance the Casimir amplitude \( c_1 = f(\Omega) \). Note that not all negative values of \( m^2 \) enhance the Casimir amplitude. As shown in Table \ref{wedge Casimir 3d}, the maximum Casimir amplitude is equivalent to that without scalars for \( m^2 = -0.3 \) and \( m^2 = -0.4 \). Note also that we retain more decimal places in our numerical computations; for simplicity, only four decimal places are listed in the tables.

In summary, we have numerically confirmed that, similar to the case of the strip, brane-localized scalar fields decrease the Casimir amplitude \( f(\Omega) \) of a wedge when \( m^2 =0 \) but can increase it for \( m^2 < 0 \). For \( m^2 = 0 \), we provide an analytical proof that the minimal tension limit of \( f(\Omega) \) is independent of the massless brane-localized scalar field. Consequently, it satisfies the holographic bound of the wedge Casimir effect for BCFTs. Finally, we note that the no-hair theorem for \( m^2 > 0 \) and \( \lambda = 0 \) still applies to a wedge.

\section{Holographic repulsive Casimir force}

The previous sections explored the attractive Casimir effect. In this section, we will examine the repulsive Casimir force. For simplicity, we will focus on strips with massless brane-localized scalars, as generalizations to wedges and massive scalars are straightforward. According to \cite{Bachas:2006ti, Diatlyk:2024qpr}, for a flat strip, the Casimir force becomes repulsive only when mixed boundary conditions are imposed on the two boundaries. As discussed in Section 3.1, the massless brane-localized scalar takes on different values at each boundary of the strip. This results in a holographic realization of the mixed boundary condition. 

 \begin{figure}[htbp]
  \centering
\includegraphics[width=1\textwidth]{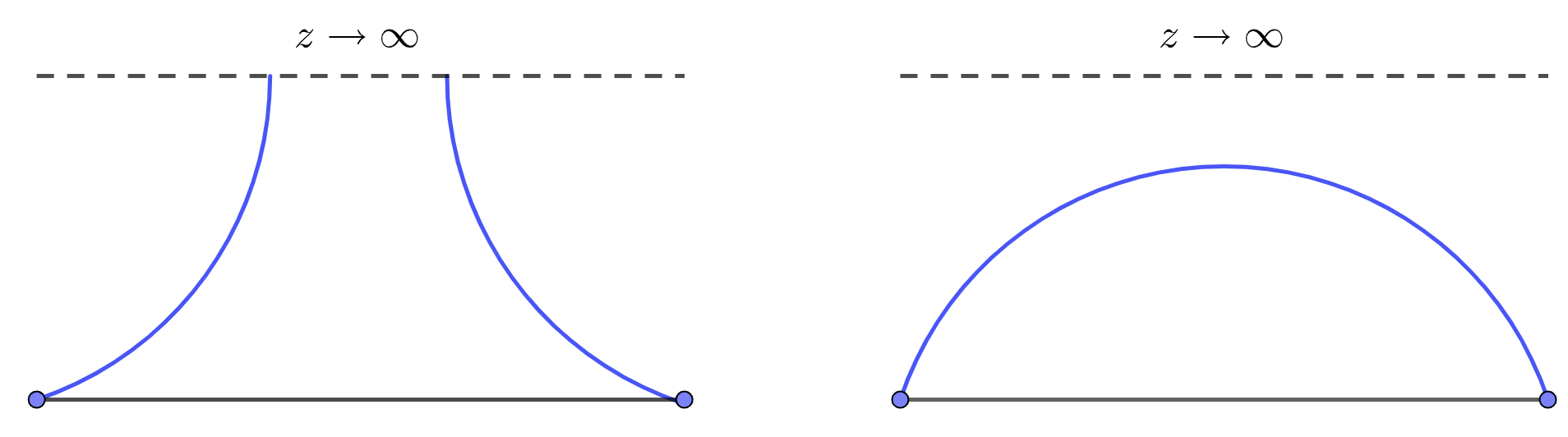}  
\caption{Disconnected EOW branes without a brane-localized scalar (Left) and connected EOW branes with a brane-localized scalar (Right) are shown for the case of the repulsive Casimir effect, with \(T_{nn} \sim -c_2 > 0\). The EOW branes and singularities are indicated by blue curves and dotted lines, respectively. In the left figure, which lacks a brane-localized scalar, the EOW branes are disconnected, and a singularity is visible as \(z \to \infty\). This scenario violates the cosmic censorship hypothesis and is therefore not permissible. In contrast, the right figure shows EOW branes that are connected because of a brane-localized scalar, with the singularity hidden behind the EOW brane. This configuration is consistent with the cosmic censorship hypothesis and indicates a repulsive Casimir effect for mixed boundary conditions.} 
 \label{scalar and repulsive Casimir}
\end{figure}

First, we explain the physical reasoning behind how a brane-localized scalar field can generate a repulsive Casimir force. For the AdS soliton, characterized by \( f(z) = 1 - c_2 z^d \), the normal-normal component of the stress tensor is given by:
\begin{align}\label{sect 5: Tnn} 
T_{nn}=-(d-1) c_2. 
\end{align} 
To achieve a repulsive Casimir force, we need to select \( c_2 < 0 \). However, this choice leads to a naked singularity as \( z \to \infty \) \cite{Huang:2026liq}. It is a curvature singularity for $d\ge 3$, while a conical singularity for $d=2$. Specifically, since \( f(z) = 1 - c_2 z^d > 0 \) for \( c_2 < 0 \), there is no horizon that satisfies \( f(z_h) = 0 \), meaning the singularity at \( z \to \infty \) remains visible. Without a brane-localized scalar field, the turning-point condition (\ref{sect 2: zmax}) with $q=0$ cannot be satisfied:
\begin{equation}\label{sect 5: turn point }
\sqrt{ f(z_{\text{max}})}\ne -\tanh(\rho). 
\end{equation}
Clearly, we have $\sqrt{ f(z_{\text{max}})}=\sqrt{ 1-c_2 z^d_{\text{max}}}>1>-\tanh(\rho)$ for $c_2<0$ and finite $\rho$. As a result, the EOW branes are disconnected in this case. The naked singularity is then part of the gravitational dual of a strip, which violates the cosmic censorship hypothesis \cite{Witten:1999xp, Galloway:1999br}. This situation is illustrated in Fig. \ref{scalar and repulsive Casimir} (Left). For this reason, \cite{Huang:2026liq} contends that the cosmic censorship hypothesis negates the possibility of a repulsive Casimir force when the same boundary conditions are applied to both boundaries of the strip. This assertion aligns with findings in field theory \cite{Bachas:2006ti, Diatlyk:2024qpr}.

With mixed boundary conditions, the Casimir force can be repulsive in principle \cite{Bachas:2006ti, Diatlyk:2024qpr}. A straightforward example of this is a \( (1+1) \)-dimensional massless free scalar field subjected to mixed Dirichlet-Neumann boundary conditions, which produces a positive Casimir pressure given by:
\begin{align}\label{sect 5: Tnn free scalar} 
T_{nn}=\frac{\pi}{48 L^2}.
\end{align} 
In our holographic model, a massless brane-localized scalar field corresponds to mixed boundary conditions. According to the holographic g-theorem \cite{Takayanagi:2011zk,Fujita:2011fp}, brane-localized matter fields that satisfy the null energy condition can cause the EOW brane to bend inward. Consequently, as illustrated in Fig. \ref{scalar and repulsive Casimir} (Right), the brane-localized scalar field transforms the EOW brane from a disconnected configuration to a connected one. We can verify that the turning-point condition (\ref{sect 2: zmax}) can indeed be satisfied for \( q\ne 0 \). This process effectively hides the singularity behind the EOW brane, in accordance with the cosmic censorship hypothesis.

Let us proceed to examine the precise gravity dual of a repulsive Casimir force. Note that the mixed boundary condition is a necessary but not sufficient condition for generating a repulsive Casimir force. As discussed in Section 3, a massless brane-localized scalar reduces the Casimir amplitude \(\kappa_1 = c_2 L^d\), but it can still yield an attractive Casimir force. To achieve a repulsive Casimir force, we must have a sufficiently large scalar difference \(\Delta \phi\), which characterizes the extent of the mixed boundary condition. For simplicity, we will focus on a scenario with negative brane tension (\(T < 0\)) and a well-defined scalar \(\sigma = 1\).
From (\ref{sect 2: dphi}) and (\ref{sect 2: dS}), we can derive expressions for the scalar difference and strip width:
\begin{equation}\label{sect 5: dphi}
\Delta \phi=2\int_0^{z_{\max}} dz\frac{2 (d-1) q z^{d-2}}{\sqrt{4 (d-1)^2 f(z)-\left(q^2  z^{2 d-2}-2 T\right)^2}},
\end{equation}
and 
\begin{equation}\label{sect 5: dS}
L=2\int_0^{z_{\max}} dz\frac{q^2  z^{2 d-2}-2 T}{f(z) \sqrt{4 (d-1)^2 f(z)-\left(q^2  z^{2 d-2}-2 T\right)^2}}.
\end{equation}
The turning point, where \(S'(z_{\text{max}}) = \infty\), gives us the following relation:
\begin{equation}\label{sect 5: zmax}
q^2=z_{\text{max}}^{2-2d}\Big(2 (d-1)\sqrt{ f(z_{\text{max}})}+2T \Big),
\end{equation}
where $f(z_{\text{max}})=1-c_2 z_{\text{max}}^d$, and $T=(d-1)\tanh(\rho)$.

\begin{figure}[htbp]
  \centering
\includegraphics[width=0.6\textwidth]{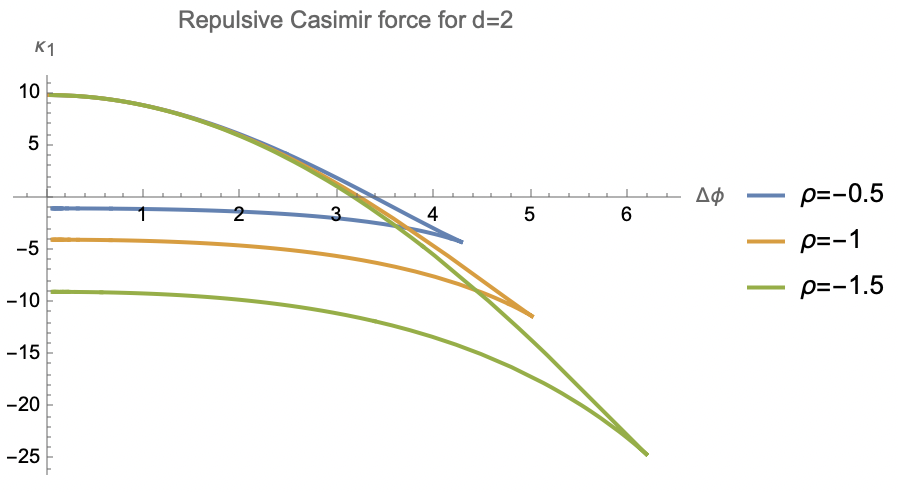} \includegraphics[width=0.6\textwidth]{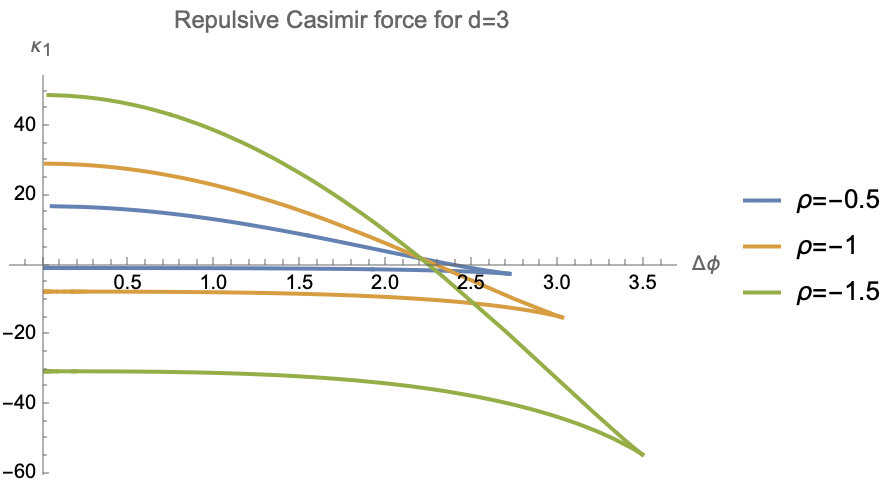} \includegraphics[width=0.6\textwidth]{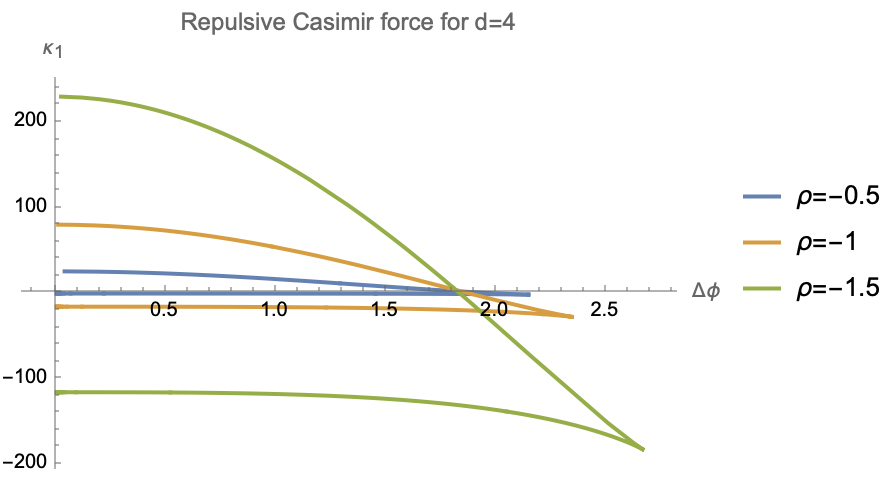}
\caption{Function $\kappa_1(\Delta \phi)$ for $d=2,3,4$ and $\rho=-0.5,-1,-1.5$. Note that each value of \(\Delta \phi\) corresponds to two values of \(\kappa_1\). We select the larger of the two \(\kappa_1\) values, as it results in a smaller Casimir energy \(-\kappa_1 A / L^{d-1}\). The figures indicate that for sufficiently large scalar difference \(\Delta \phi\), the Casimir amplitude \(\kappa_1\) can become negative, which suggests a repulsive Casimir force for mixed boundary conditions.} 
 \label{repulsiveCasimir}
\end{figure}

We fix the brane tension \(T\) and the strip width \(L=1 \).
Consequently, equations (\ref{sect 5: dS}) and (\ref{sect 5: zmax}) allow us to establish a relationship between \(z_{\text{max}}\) and \(c_2\). By substituting \(z_{\text{max}}(c_2)\) and (\ref{sect 5: zmax}) into (\ref{sect 5: dphi}), we can determine the relationship between \(\Delta \phi\) and \(c_2\). Note that for \(L = 1\), we have \(\kappa_1 = c_2 L^d = c_2\). We then numerically derive the relationship between the Casimir amplitude \(\kappa_1\) and the magnitude of the mixed boundary condition \(\Delta \phi\). As illustrated in Fig. \ref{repulsiveCasimir}, for a sufficiently large scalar difference \(\Delta \phi\), the Casimir amplitude \(\kappa_1\) can become negative, indicating a repulsive Casimir force.

As shown in Fig. \ref{repulsiveCasimir}, one $\Delta \phi$ corresponds to two values of $\kappa_1$. We select the larger value of $\kappa_1$ because it results in a smaller Casimir energy, expressed as $-\kappa_1 A /L^{d-1}$. Let us clarify this further. Note that the holographic bulk energy density of BCFTs is given by
\begin{align}\label{sect 5: Casimir energy density} 
T_{tt}=-c_2.
\end{align} 
In contrast, the holographic boundary energy density of BCFTs is zero, i.e., $\tau_{tt} = 0$. This occurs because the boundary's intrinsic and extrinsic curvatures both vanish for a flat strip. Consequently, the total Casimir energy can be expressed as:
\begin{align}\label{sect 5: Casimir energy} 
W=T_{tt}A L=-c_2A L=-\kappa_1 A /L^{d-1},
\end{align}  
where $A$ represents the transverse area and $L$ denotes the width of the strip. Therefore, we choose the curves with larger values of $\kappa_1$ in Fig. \ref{repulsiveCasimir}, as they correspond to a lower free energy. The free energy equals the Casimir energy at zero temperature.

In summary, we have demonstrated that a sufficiently large brane-localized scalar field can produce a repulsive Casimir force. Furthermore, we have argued that this holographic repulsive Casimir force is consistent with the cosmic censorship hypothesis.

\section{Conclusions and Discussions}

This paper investigates the holographic Casimir effect under mixed boundary conditions in CFTs and non-CFTs by introducing a scalar field on the EOW brane. The massless brane-localized scalar corresponds to an exact boundary marginal deformation, yielding a new family of BCFTs. We provide analytical proofs for negative tensions and numerical verifications for general tensions that this new BCFT adheres to the holographic bound of the Casimir effect for BCFTs \cite{Miao:2024gcq, Miao:2025utb}. In the case of massive brane-localized scalars, a boundary-irrelevant deformation occurs when the mass squared \( m^2 > 0 \), while a boundary-relevant deformation arises when \( m^2 < 0 \). We prove a no-hair theorem for positive mass squared \( m^2 > 0 \) and zero boundary source \( \lambda = 0 \). This theorem implies that the massive brane-localized scalar cannot support a connected EOW brane linking the two boundaries. Consequently, we focus on the relevant deformation with \( m^2 < 0 \), which generally transforms a CFT into a non-CFT. For non-CFTs, the norm of the displacement operator \( C_D \) is no longer constant, leading to an open question regarding how to extend the bound on the ratio \( -\kappa_1/C_D \) to non-CFTs. Therefore, we center our attention on the Casimir amplitude \( \kappa_1 \) for non-CFTs. For dimensions \( d \ge 3 \), we find that massive brane-localized scalars can increase the Casimir amplitude for \( m^2 < 0 \). This suggests that we can enhance the Casimir effect by implementing a boundary-relevant deformation. The \( d = 2 \) case is distinct, and by applying the holographic g-theorem, we demonstrate that the brane-localized scalar decreases the Casimir amplitude. For \( d = 2 \), a natural choice for \( C_D \) is the bulk central charge \( c \), which remains independent of boundary deformations. In this way, we establish the holographic bound of \( -\kappa_1/C_D\sim-\kappa_1/c \) for general two-dimensional brane-localized scalar fields. These conclusions are valid for both the holographic strip and the holographic wedge.

Interestingly, we find that a sufficiently large brane-localized scalar field can generate a repulsive Casimir force. From a field-theoretic perspective, a repulsive Casimir force can be produced when we impose mixed boundary conditions. In this case, the brane-localized scalar takes different values at the two boundaries of the strip, which corresponds to a mixed boundary condition capable of producing a repulsive Casimir force. From a holographic viewpoint, the repulsive Casimir effect faces the naked-singularity problem associated with identical boundary conditions without brane-localized scalars. However, in our case of mixed boundary conditions with brane-localized scalars, the scalars bend the EOW brane inward, transforming the configuration from disconnected to connected. Consequently, the singularity becomes hidden behind the EOW brane, aligning with the cosmic censorship hypothesis. 

For simplicity, this paper focuses on brane-localized scalar fields. It would be interesting to expand this discussion to include bulk matter fields. Additionally, clarifying how to generalize the definition of the ratio \( -\kappa_1/C_D \) for non-CFTs remains a goal for future research.

\acknowledgments

We acknowledge the supports from National Natural Science Foundation of China (NSFC) grant (No.12275366).


\appendix

\section{Monotonicity of the strip width}

This appendix demonstrates that a massless brane-localized scalar field reduces the strip width \( L \) when the brane tension is negative.

We introduce the following normalized parameters:
\begin{align}\label{app: b c n}
    b=\frac{T}{d-1},\qquad c=\frac{q^2}{2(d-1)},\qquad n=2d-2.
\end{align}
For \( T < 0 \) and \( q > 0 \), it follows that \( -1 < b < 0 \) and \( c > 0 \).   Since $T<q^2/2$, only the first branch of the strip width occurs, namely
\begin{equation}   \label{app: L-original}
    L=2\int_0^{z_{\text{m}}}
    \frac{c z^n-b}
    {(1-z^d)\sqrt{1-z^d-(c z^n-b)^2}}\,dz .
\end{equation}
The turning point $z_{\text{m}}=z_{\max} \in(0,1)$ is uniquely determined by
\begin{equation}    \label{app: turning}
    c z_{\text{m}}^n-b=\sqrt{1-z_{\text{m}}^d}.
\end{equation}
Indeed, the left-hand side of \eqref{app: turning} is strictly increasing and the right-hand side is strictly decreasing.

Set \( z = z_{\text{m}} u \), \( B = b + \sqrt{1 - z_{\text{m}}^d} = cz_{\text{m}}^n > 0 \), and define the following functions:
\begin{align} \label{app: F G Delta}
    F(u)&=1-z_{\text{m}}^d u^d,\\
    G(u)&=B u^n-b,\\
    \Delta(u)&=F(u)-G(u)^2.
\end{align}
Then, the strip width (\ref{app: L-original}) can be expressed as:
\begin{equation} \label{app: L}
    L(z_{\text{m}})=2z_{\text{m}}\int_0^1\frac{G(u)}{F(u)\sqrt{\Delta(u)}}\,du,
\end{equation}
where the turning-point relation indicates that \( \Delta(1) = 0 \).
We also consider the following two useful formulas:
\begin{align}\label{app: key formula 1}
\frac{d}{d z_{\text{m}}} \Big( \frac{ z_{\text{m}}G}{F\sqrt{\Delta}}\Big)= \frac{u^{2d-2} R}{\sqrt{1-z_{\text{m}}^d}\Delta^{\frac{3}{2}}}+ \frac{d}{du}\Big( \frac{ Q}{F\sqrt{\Delta}} \Big),
\end{align}
and \begin{align}\label{app: key formula 2}
\Big( \frac{ Q}{F\sqrt{\Delta}} \Big) \Big|^1_0=0,
\end{align}
where $R$ and $Q$ are given by
\begin{align}\label{app: R}
    R=\frac{3d-2}{2}z_{\text{m}}^d(1-u^d)+B(1-u^{2d-2})\left[B(1+u^{2d-2})-2d b\right],
\end{align}
and 
\begin{align}\label{app: Q}
    Q=\frac{-u}{\sqrt{1-z_{\text{m}}^d}} \Big( b \sqrt{1-z_{\text{m}}^d} (1-u^{2d-2})+z_{\text{m}}^d u^{2d-2} (1-u^d) \Big). 
\end{align}

Taking the derivative of (\ref{app: L}) with respect to $z_{\text{m}}$ and applying formulas (\ref{app: key formula 1}, \ref{app: key formula 2}), we obtain
\begin{equation}    \label{app: derivative-identity}
    \frac{1}{2}\frac{dL}{dz_{\text{m}}}
    =\frac{1}{\sqrt{1-z_{\text{m}}^d}}\int_0^1
      \frac{u^{2d-2} R(u)}{\Delta(u)^{3/2}}\,du.
\end{equation}
For $0\leq u < 1$, $b<0$ and $B>0$, we have $R(u)> 0$.  Consequently, \eqref{app: derivative-identity} implies that
\begin{equation}   \label{app: LP-positive}
    \frac{dL}{dz_{\text{m}}}>0.
\end{equation}

It remains to compare $z_{\text{m}}$ with the scalar parameter $c=q^2/(2(d-1))$.  From the turning-point equation (\ref{app: turning}),
\begin{equation}
    c=\frac{b+\sqrt{1-z_{\text{m}}^d}}{z_{\text{m}}^{2d-2}}.
\end{equation}
we derive     
\begin{equation} \label{app: cP-negative}
\frac{dc}{dz_{\text{m}}}= \frac{1}{2} z_m^{1-2 d} \left(-4 (d-1) \left(b+\sqrt{1-z_m^d}\right)-\frac{d z_m^d}{\sqrt{1-z_m^d}}\right)<0.
\end{equation}
On the other hand, $c=q^2/(2(d-1))$ with $q>0$ gives
\begin{equation} \label{app: dc dq}
    \frac{dc}{dq}=\frac{q}{(d-1)}>0.
 \end{equation}
Equations \eqref{app: LP-positive} and \eqref{app: cP-negative} now yield
\begin{equation} \label{app: key result}
    \frac{dL}{dq}
      =\frac{dL}{dz_{\text{m}}}\frac{dz_{\text{m}}}{dc}\frac{dc}{dq}<0.
 \end{equation}
This proves the claim that, for negative brane tension with \( b = \tanh(\rho) < 0 \), the massless brane-localized scalar decreases the strip width.

{\bf{Remark on positive tension.}}

When \( T > 0 \), the factor 
\[ B(1 + u^{2d - 2}) - 2db \] 
in (\ref{app: R}) may not be positive, which means the previous argument is no longer applicable. However, high-precision numerical integration across both branches consistently indicates that
 \[ \frac{dL}{dq} < 0 \] 
 for the sampled physical region where \( L \geq 0 \). This provides numerical evidence supporting the positive-tension monotonicity of $L$. We will leave the analytic proof for the case when \( T > 0 \) for future work.

\end{document}